\documentclass[aps,prd,preprintnumbers,superscriptaddress,nofootinbib,notitlepage]{revtex4-2}
\usepackage[pdftex]{graphicx}
\usepackage{bm,latexsym,amsmath,amssymb,amsfonts,mathrsfs}
\usepackage{mathtools}
\usepackage{color}
\allowdisplaybreaks[1]
\usepackage[pdftex,colorlinks=true,linkcolor=blue,citecolor=cyan]{hyperref}

\begin{document}
\title{Vector-tensor theories in metric-affine geometry}
\author{Tact~Ikeda}
\email[Email: ]{tact@rikkyo.ac.jp}
\affiliation{Department of Physics, Rikkyo University, Toshima, Tokyo 171-8501, Japan
}
\begin{abstract}
We investigate ghost-free vector-tensor theories in metric-affine geometry.
In all of our analysis, we start with the Lagrangian containing up to quadratic terms of first-order derivatives of a vector field.
To obtain ghost-free vector-tensor theories efficiently, we consider two options; the theories satisfy the torsionless condition or have the projective symmetry.
We first explore the vector-tensor theories under the former condition.
We then investigate the projective-invariant vector-tensor theories in metric-affine geometry.
To systematically construct a projective-invariant Lagrangian, we use two different approaches.
First, we construct a Lagrangian by contracting the epsilon tensor.
Second, we construct a Lagrangian by use of projective-invariant combinations.
We find that to obtain a ghost-free Lagrangian in metric-affine geometry, imposing the projective invariance would be more useful than imposing the torsionless condition.
However, we also prove that the projective invariance (or the torsionless condition) alone is insufficient for vector-tensor theories in metric-affine geometry to be ghost-free.
\end{abstract}
\preprint{RUP-23-24}
\maketitle
\section{Introduction}

General relativity (GR) is the successful, modern, and standard theory of gravity.
Nevertheless, modified theories of gravity are actively investigated as well for several reasons.
For instance, GR is considered as a low-energy effective theory, and hence might not work well in a strong gravity regime (e.g., the early Universe and the vicinity of black holes).
Therefore, it is worth constructing a unified framework of gravity, which can be used even in a high-energy system and has GR as a certain limit.
In addition, modified theories of gravity are useful as a comparison for GR, allowing us to test gravity (see, e.g., Refs.~\cite{Koyama:2015vza,Ferreira:2019xrr,Arai:2022ilw} for reviews).
Indeed, having achieved the direct detection of gravitational waves~\cite{LIGOScientific:2016aoc} and imaging of black hole shadows~\cite{EventHorizonTelescope:2019dse} in recent years, testing gravity in a strong gravity regime increasingly become practical.
Also, GR may not resolve various enigmas of the Universe.
For instance, the mechanism of the late-time accelerated expansion of the Universe is still unclear.
Thus, the modification of GR at large distances is actively studied as an alternative to dark energy.
Finally, the study of modified gravity theory helps us to deeply understand gravity, e.g., the specificity of GR.
For instance, there exist the following open questions: how special a massless graviton is, how special gravity in four dimensions is, and how special gravity in (pseudo)-Riemannian geometry is.
These issues would be resolved by investigating massive gravity, gravity in higher (or lower) dimensions, and gravity in general geometry, respectively.
Therefore, it is also worth trying to capture the nature of gravity through the study of modified gravity.

As is well-known,
GR is described on curved spacetime, and then the appearance of a geometrical object (i.e., the curvature tensor) is interpreted as the contribution of gravity.
For instance, when spacetime is only slightly curved, i.e., the value of the geometrical object is very small, GR reproduces Newtonian gravity.
Usually, ``curved spacetime'' denotes geometry on (pseudo-)Riemannian manifold, and GR (and most modified gravity) is also described in such geometry.
However, in general, curved spacetime is not only (pseudo-)Riemannian geometry.
In particular, metric-affine geometry (see, e.g., Refs.~\cite{Hehl:1994ue,Kleyn:2004yj,Blagojevic:2013xpa} for a review) is actively studied, which is an extension of (pseudo-)Riemannian geometry and there exist additional geometrical objects (i.e., the torsion and non-metricity tensor).
Then, geometrical objects are described by two variables, that is, a metric and a connection, whereas, in (pseudo-)Riemannian geometry, the geometrical object is completely governed by a metric alone.
Therefore, the additional geometrical objects, i.e., the torsion tensor and the non-metricity tensor give us a new method for modification of gravity.
Indeed the Einstein-Hilbert action in the metric-affine formalism is effectively equivalent to GR in vacuum~\cite{Giachetta:1997tst,Sotiriou:2009xt,Dadhich:2012htv},
but there is no guarantee that the equivalence holds even in the context of modified gravity.
For instance, the equivalence between metric-affine and (pseudo-)Riemannian geometry does not hold once one introduces higher-order curvature terms (e.g., Refs.~\cite{Sotiriou:2006qn,Sotiriou:2009xt,Janssen:2019doc,Janssen:2019uao,BeltranJimenez:2019acz,BeltranJimenez:2020sqf}) or extra fields (e.g., Refs.~\cite{Aoki:2018lwx,Helpin:2019kcq,Helpin:2019vrv}).
In addition, there seems to be no reason why only (pseudo-)Riemannian geometry is used.
Thus, in particular, we would like to investigate (modified) gravity on metric-affine geometry.

Aside from metric-affine geometry, let us focus on the way of introducing extra fields to extend GR.
If one introduces new gravitational degree(s) of freedom (DOFs) in addition to the tensor DOFs by adding extra fields, one may be faced with higher-order derivative terms on the action.
However, such higher-derivative theories suffer from the Ostrogradsky instability in general~\cite{Woodard:2015zca,Motohashi:2020psc,Aoki:2020gfv}.
To evade this unwanted ghost DOF(s), higher-derivative theories should be degenerate.
In a nutshell, degenerate higher-order theories yield less number of DOFs than expected from the order of derivatives in the theories so that there exist no extra unwanted DOFs caused by higher derivatives~\cite{Motohashi:2014opa,Motohashi:2016ftl,Klein:2016aiq,Motohashi:2017eya,Motohashi:2018pxg}.
For instance, the Horndeski theory~\cite{Horndeski:1974wa,Deffayet:2011gz,Kobayashi:2011nu}, which is known as the most general degenerate (single-field) scalar-tensor theory equipping second-order Euler-Lagrange equations, is trivially degenerate and hence free from the Ostrogradsky instability.
Furthermore, the unified framework of a large family of higher-derivative scalar-tensor theories with a single scalar and two tensor DOFs is Degenerate higher-order scalar-tensor (DHOST) theories~\cite{Langlois:2015cwa,Crisostomi:2016czh,BenAchour:2016fzp} (see Refs.~\cite{Langlois:2018dxi,Kobayashi:2019hrl} for reviews).
DHOST theories include, e.g., the so-called ``beyond Horndeski'' or
GLPV theories~\cite{Gleyzes:2014dya,Gleyzes:2014qga} as specific cases.
So far, we have discussed only degenerate scalar-tensor theories.
However, there is also another way of modification of gravity by adding vector field DOFs instead of scalar field, i.e., vector-tensor theories.
In the context of massless vector-tensor theories, there is a no-go theorem that if one imposes the $U(1)$ symmetry on theory in flat spacetime, Galileon-like terms are prohibited~\cite{Deffayet:2013tca}.
Note that, in curved spacetime, the vector Horndeski theory~\cite{Horndeski:1976gi} exists as a massless vector-tensor theory.
On the other hand, massive vector-tensor theories are actively studied, and a prime example is generalized Proca theory~\cite{Heisenberg:2014rta}.
Generalized Proca theory is free from the Ostrogradsky instability because it is trivially degenerate as in the case of Horndeski theory.
In addition, like DHOST theories, there are ``Extended vector-tensor theories~\cite{Kimura:2016rzw}'' as a framework for degenerate vector-tensor theories, which includes many vector-tensor theories, such as generalized Proca theory and ``beyond generalized Proca theory~\cite{Heisenberg:2016eld}.''

In this paper, we will investigate vector-tensor theories in metric-affine geometry.
As mentioned above, modified gravity in metric-affine geometry would not coincide with one in (pseudo-)Riemannian geometry in general.
Furthermore, the extension to metric-affine geometry is not unique in principle, and most of them would suffer from Ostrogradsky instability.
We thus explore the proper extensions of vector-tensor theories in metric-affine geometry and how to construct them by considering the Lagrangian constructed by up to quadratic terms of first-order derivatives of a vector field as a first step.

This paper is organized as follows.
In the next section, we overview the major properties of metric-affine geometry and scalar-tensor theories in metric-affine geometry.
We introduce geometric quantities defined on metric-affine geometry and derive some formulae.
Especially, we refer to the projective transformation and its importance.
In the last of this section, we review previous studies on the scalar-tensor theories in metric-affine geometry.
Then, we show that imposing the projective invariance (or the torsionless condition) on the theory is useful to evade the Ostrogradsky instability.
In Sec.~\ref{sec:VTInPalatini}, we focus on torsionless vector-tensor theories in metric-affine geometry and discuss the degeneracy conditions for this system.
In Sec.~\ref{sec:VTInMA}, we focus on projective-invariant vector-tensor theories in metric-affine geometry.
We introduce two options to obtain a projective-invariant Lagrangian.
Also, we will show the degeneracy conditions in each of those cases.
Finally, we draw our conclusions in Sec.~\ref{sec:concl}

\section{Preliminaries}

\subsection{Basis quantities in metric-affine geometry and notations}

In this section, we introduce geometrical objects in metric-affine geometry, which we will use in the following calculations.
In metric-affine geometry, we use a general affine connection, which does not necessarily coincide with the Levi-Civita connection.
We therefore introduce the two kinds of affine connections as follows:
\begin{align}
    \nabla_{a}A_{b}&\coloneqq\partial_{a}A_{b}-\Gamma^{c}{}_{ab}A_{c},\\
    D_{a}A_{b}&\coloneqq\partial_{a}A_{b}-\{{}^{\,c}_{ab}\}A_{c},
\end{align}
where $\Gamma^{c}{}_{ab}$ is a general connection and
\begin{align}
    \{{}^{\,a}_{bc}\}\coloneqq\frac{1}{2}g^{ad}(\partial_{b}g_{dc}+\partial_{c}g_{bd}-\partial_{d}g_{bc}),
\end{align}
is the Levi-Civita connection.
We can then define the two types of curvature tensors as
\begin{align}
    R_{abc}{}^{d}(\Gamma)&\coloneqq\Gamma^{e}{}_{ac}\Gamma^{d}{}_{be}-\Gamma^{e}{}_{bc}\Gamma^{d}{}_{ae}-\partial_{a}\Gamma^{d}{}_{bc}+\partial_{b}\Gamma^{d}{}_{ac}, \\
    \overset{g}{R}{}_{abc}{}^{d}(g)&\coloneqq\{{}^{\,e}_{ac}\}\{{}^{\,d}_{be}\}-\{{}^{\,e}_{bc}\}\{{}^{\,d}_{ae}\}-\partial_{a}\{{}^{\,d}_{bc}\}+\partial_{b}\{{}^{\,d}_{ac}\}. \label{eq:RiemannLeviCivita}
\end{align}
The curvature tensor naturally induces its trace, the Ricci tensor and Ricci scalar respectively, as\footnote{%
Strictly speaking, there exist other independent trace components of the curvature tensor, i.e., the homothetic tensor
\begin{align}
	H_{ab}(\Gamma)=R_{bac}{}^c,
\end{align}
and the co-Ricci curvature
\begin{align}
	P^a{}_d(\Gamma,g)=g^{bc}R_{dcb}{}^{a}.
\end{align}
However, in this work, these trace components will not be used since they will not appear explicitly in our analysis.}
\begin{align}
    R_{ab}(\Gamma)&\coloneqq R_{adb}{}^{d},  \label{eq:RicciTensor}\\
    R(\Gamma,g)&\coloneqq g^{ab}R_{ab}.\label{eq:RicciScalar}
\end{align}
When we use the curvature tensors built out of the Levi-Civita tensor rather than a general connection, we put $g$ on top of them as in Eq.~\eqref{eq:RiemannLeviCivita}.
In four dimensions, the Einstein tensor $G^{ab}$ in metric-affine geometry can be defined by
\begin{align}
    G^{a b c d}&\coloneqq\frac{1}{4}\epsilon^{aba'b'}\epsilon^{cdc'd'}R_{a'b'c'd'},
    \label{et01}
    \\
    G^{ab}&\coloneqq G_{c}{}^{acb}
    \label{et02},
\end{align}
where the epsilon tensor (the completely antisymmetric tensor) is defined using the generalized Kronecker delta as
\begin{align}
    \epsilon_{abcd}\coloneqq\sqrt{|g|}\delta^{0123}_{abcd} =4! \sqrt{|g|}\delta^0_{[a} \delta^1_b \delta^2_c \delta^3_{d]},
\end{align}
with $g\coloneqq\det [g_{ab}]$.

Let us introduce the torsion and non-metricity tensors defined as follows:
\begin{align}
    T^{a}{}_{bc}&\coloneqq \Gamma^{a}{}_{bc}-\Gamma^{a}{}_{cb},  \\
    Q_{abc}&\coloneqq -\nabla_{a}g_{bc}.
\end{align}
Note that for the Levi-Civita connection, we have $\{{}^{\,a}_{bc}\}-\{{}^{\,a}_{cb}\}=0$ and $D_a g_{bc}=0$.
The affine connection can be decomposed into a sum of the Levi-Civita connection and a post-Riemannian term (the distortion tensor) as
\begin{align}
    \Gamma^a{}_{bc}=\{{}^{\,a}_{bc}\}+\kappa^a{}_{bc}.
\end{align}
Here, the distortion tensor $\kappa^a{}_{bc}$ is given explicitly in terms of the torsion and non-metricity tensors as
\begin{align}
    \kappa^{a}{}_{bc}
    &\coloneqq \frac{1}{2}(Q_{bc}{}^a+Q_{cb}{}^a-Q^a{}_{bc})-\frac{1}{2}(T_{bc}{}^a+T_{cb}{}^a-T^a{}_{bc}).
\end{align}
Using the distortion tensor, one can write the curvature tensor, the torsion tensor, and the non-metricity tensor as
\begin{align}
    R_{abc}{}^{d}&=\overset{g}{R}{}_{abc}{}^{d}+\kappa^{e}{}_{ac}\kappa^{d}{}_{be}-\kappa^{e}{}_{bc}\kappa^{d}{}_{ae}-D_{a}\kappa^{d}{}_{bc}+D_{b}\kappa^{d}{}_{ac}, \\
    T^{a}{}_{bc}&=\kappa^{a}{}_{bc}-\kappa^{a}{}_{cb}, \\
    Q_{abc}&=\kappa_{bac}+\kappa_{cab}.
\end{align}

Let $L$ be our gravitational Lagrangian.
In vacuum we have the Euler-Lagrange equations for $\Gamma^a{}_{bc}$,
\begin{align}
\mathcal{P}_a{}^{bc}&\coloneqq \frac{1}{\sqrt{-g}}\frac{\delta(\sqrt{-g}L)}{\delta \Gamma^a{}_{bc}}=\frac{1}{\sqrt{-g}}\frac{\delta(\sqrt{-g}L)}{\delta \kappa^a{}_{bc}}
=0.\label{eq:kappa-gamma-L}
\end{align}
We will consider the case where Eq.~\eqref{eq:kappa-gamma-L} gives constraint equations for $\kappa^a{}_{bc}$ and can be solved explicitly to obtain $\kappa^a{}_{bc}$ in terms of the other fields.
Note that the variation with respect to $\Gamma^a{}_{bc}$ leads to the same tensor $\mathcal{P}_a{}^{bc}$ as that with respect to $\kappa^a{}_{bc}$.
This fact is useful in actual calculations. It should be emphasized, however, that the independent variables are the metric and the connection, but not the metric and the distortion tensor.

\subsection{Projective transformation}

Let us consider the following transformation of the affine connection with a vector field $\xi^a$:
\begin{align}
    \Gamma^{a}{}_{bc}\to\hat \Gamma^{a}{}_{bc}=\Gamma^{a}{}_{bc}+\delta^a_c\xi_b. \label{eq:projectiveT}
\end{align}
This is called the projective transformation.
This transformation retains the form of the auto-parallel equation, which is well-known as the geodesic equation in (pseudo-)Riemannian geometry.
Under the projective transformation, the curvature tensors transform as
\begin{align}
    R_{abc}{}^d\to&\; \hat R_{abc}{}^d=R_{abc}{}^d-\mathcal{F}_{ab}\delta^d_c, \\
    R_{ab}\to&\; \hat R_{ab}=R_{ab}-\mathcal{F}_{ab}, \\
    R\to&\; \hat R=R,\label{Eq:prjtrR}
\end{align}
where
\begin{align}
    \mathcal{F}_{ab}\coloneqq\partial_a \xi_b-\partial_b \xi_a.
\end{align}
Equation~\eqref{Eq:prjtrR} shows that the Einstein-Hilbert action in the metric-affine formalism is projective invariant.

Several studies have pointed out that the projective invariance is crucial for a theory of gravity to be free from ghost instabilities (see, e.g., Refs.~\cite{BeltranJimenez:2019acz,BeltranJimenez:2020sqf,Aoki:2018lwx,Helpin:2019vrv}).
In the case of scalar-tensor theories, this point will be reviewed and discussed in more detail in the next subsection.

\subsection{Scalar-tensor theories in the Palatini and the metric-affine formalisms}\label{sec:STInMA}

In this section, we consider the Horndeski theory~\cite{Horndeski:1974wa,Deffayet:2011gz,Kobayashi:2011nu} as an example to grasp the general idea of how ``healthy'' (i.e., ghost-free) theories of gravity can be obtained in metric-affine geometry (i.e., in the metric-affine formalism).
The goal is to extend the Horndeski theory in (pseudo-)Riemannian geometry (i.e., in the metric formalism) to its metric-affine counterpart.
Here, there are multiple ways to extend the theory, some of which yield healthy theories and some do not.
We therefore wish to explore an efficient way to obtain healthy theories.

Hereafter, for simplicity, we focus on a certain subset of the Horndeski family and assume that the Lagrangian is at most quadratic in the second derivative of the scalar field $\phi$.
The reason for this simplification is that the Lagrangian containing cubic terms in the second derivative of $\phi$ leads to the Euler-Lagrange equations ${\cal P}_a{}^{bc}=0$ that involve quadratic terms in the connection and hence are difficult to solve explicitly for the connection.
Now we further comment on this simplification.
This is just a prescription for technical difficulties, not one for physical difficulties.

The first way we consider for obtaining ghost-free theories is to impose the torsionless condition,
\begin{align}
    T^a{}_{bc}=0,
\end{align}
at the level of the Lagrangian.
In this paper, we call the metric-affine formalism with this condition the \textit{Palatini formalism}.\footnote{%
There seems to be some ambiguity in the use of the terminology the \textit{Palatini formalism} in the literature.}
It is easy to see, however, that imposing the torsionless condition alone does not uniquely fix the form of Lagrangian
for the metric-affine version of the Horndeski theory.
We, therefore, choose to impose the simplest extra condition that the covariant derivative with the Levi-Civita connection, $D_a$, in the Lagrangian in the metric formalism is replaced by the torsionless covariant derivative, $\nabla_a$, to move to the metric-affine
formalism.\footnote{%
Two comments are here in order.
First, precisely speaking, replacing $D_a\to\nabla_a$ is not unique.
The proper statement is that we rewrite all second-derivative (of the scalar field) into $D^a\phi_b$, replacing $D^a\phi_b\to \nabla^a\phi_b$.
It is clear that the procedure is unique.
Second, it would be better to use different symbols for the general covariant derivative and the torsionless covariant derivative. Nevertheless, we use the same ones since it does not cause any confusion.}
This procedure has been adopted in Ref.~\cite{Helpin:2019kcq} to study the stability and cosmology of the metric-affine theory with the Lagrangian
\begin{align}
    L=G_2(\phi,\mathcal{X})+G_3(\phi,\mathcal{X})\nabla^a  \phi_a+G_4(\phi)R,
\end{align}
where $\phi_a\coloneqq\nabla_a \phi$ and $\mathcal{X}\coloneqq-\phi^a\phi_a/2$.
Furthermore, Ref.~\cite{Helpin:2019kcq} investigated the case with the more general Lagrangian
\begin{align}
    L=G_2(\phi,\mathcal{X})+G_3(\phi,\mathcal{X})\nabla^a \nabla_a \phi+G_4(\phi,\mathcal{X})R+G_{4\mathcal{X}}(\phi,\mathcal{X})[(\nabla^a\phi_a)^2-\nabla^a\phi_b\nabla^b\phi_a] \label{eq:HorndeskiInPalatini},
\end{align}
and showed that a ghost-free theory is obtained at least if $G_4$ is of the form $G_4=f(\phi)\mathcal{X}$.

The second way to obtain ghost-free theories is to impose the projective invariance on the Lagrangian.
As an example, let us see how one can impose the projective invariance in extending the so-called cubic Horndeski Lagrangian,
\begin{align}
	L_3=G_3(\phi,\mathcal{X})D^a\phi_a,\label{eq:Horn-L3}
\end{align}
to its metric-affine version $L_3^{\textrm{MA}}$. In fact, just for the purpose of obtaining a ghost-free theory, it is not necessary to require the projective invariance of $L_3^{\textrm{MA}}$.
This is, however, not true if one includes the terms quadratic in second derivatives of $\phi$. We therefore begin with this simpler example to learn the way to implement the projective invariance.
We can consider the following four building blocks to express the second derivative of $\phi$ in the metric-affine formalism~\cite{Helpin:2019vrv}:
\begin{align}
    \overset{(1)}{\Phi}{}^a{}_b=\nabla^a \phi_b, \quad
        \overset{(2)}{\Phi}{}^a{}_b=g_{bc}\nabla^a \phi^c,  \quad 
    \overset{(3)}{\Phi}{}^a{}_b=(1/4)g_{cd}\nabla^a(g^{cd}\phi_b), \quad
        \overset{(4)}{\Phi}{}^a{}_b=\nabla_c(g^{ac}\phi_b) \label{eq:TermsOfPhi}.
\end{align}
All of these terms reduce to $D^a\phi_b$ in the limit of $\kappa^a{}_{bc}\to 0$.
We look for an extension of $L_3$ that is built out of these four terms and has the projective invariance.
Note, however, that we have already restricted ourselves to a subset of infinitely many possible forms of $L_3^{\textrm{MA}}$, as one may in principle consider the Lagrangian of the form $L_3^{\textrm{MA}}=G_3D^a\phi_a+L'(\kappa^a{}_{bc}, g_{ab}, \phi;D_a)$, where $L'$ is some function satisfying $L'(0,g_{ab},\phi)=0$ so that $L_3^{\textrm{MA}}=L_3$ in the $\kappa^a{}_{bc}\to 0$ limit~\cite{Helpin:2019vrv}.
This restriction comes from a practical reason: integrating out the connection is not feasible if one works with such a general Lagrangian.

Under the projective transformation, we have 
\begin{align}
    \overset{(1)}{\Phi}{}^a{}_b\to \overset{(1)}{\Phi}{}^a{}_b
    -\xi^a\phi_b,
    \quad 
    \overset{(2)}{\Phi}{}^a{}_b\to \overset{(2)}{\Phi}{}^a{}_b
    +\xi^a\phi_b,
    \quad 
    \overset{(3)}{\Phi}{}^a{}_b\to \overset{(3)}{\Phi}{}^a{}_b
    +\xi^a\phi_b,
    \quad 
    \overset{(4)}{\Phi}{}^a{}_b\to \overset{(4)}{\Phi}{}^a{}_b
    +\xi^a\phi_b.
    \label{phi-p-tr}
\end{align}
Therefore, it is obvious that a straightforward replacement $D^a\phi_a\to \overset{(1)}{\Phi}{}^a{}_a$ does not work.
One can, however, write a projective invariant combination constructed only out of $\overset{(1)}{\Phi}{}^a{}_b$ as
\begin{align}
    H_3(\phi,\mathcal{X})\left(-
    \phi_c\phi^c\delta_a^b +\phi_a\phi^b
    \right)\overset{(1)}{\Phi}{}^a{}_b.\label{eq:H3xpp}
\end{align}
In the $\kappa^a{}_{bc}\to 0$ limit, this term reduces to $H_3\left(2\mathcal{X}D^a\phi_a+\phi_a\phi^b D^a\phi_b\right)$.
This does not apparently coincide with Eq.~\eqref{eq:Horn-L3}, but by performing an integration by parts one can recast it into the form of Eq.~\eqref{eq:Horn-L3} (plus an additional unimportant term containing the first derivative of $\phi$).
Therefore, Eq.~\eqref{eq:H3xpp} can be regarded as a projective-invariant metric-affine generalization of the cubic Horndeski Lagrangian.

Now, it is suggestive to see that Eq.~\eqref{eq:H3xpp} can also be expressed using the epsilon tensor as
\begin{align}
	\frac{H_3}{2}\epsilon^{abcd}\epsilon^{a'b'}{}_{cd}\phi_a\phi_{a'}\nabla_b\phi_{b'}. \label{eq:HorndeskiL3epsilonChange}
\end{align}
This fact hints at a convenient way to derive a projective-invariant combination even at higher order in second derivatives of $\phi$, as developed in Ref.~\cite{Aoki:2018lwx}.
The authors of Ref.~\cite{Aoki:2018lwx} have studied the metric-affine scalar-tensor theory with the Lagrangian
\begin{align}
    L&=f_0 R +f_1 G^{ab} \phi_a \phi_b+f_2+f_3\mathcal{L}_3^{\rm gal \phi}+f_4\mathcal{L}_4^{\rm gal \phi},\label{Lag:shim-aoki}
\end{align}
where $f_i$ ($i=0,1,2,3,4$) are arbitrary functions of $\phi$ and $\mathcal{X}$, and
\begin{align}
	\mathcal{L}_3^{\rm gal\phi}\coloneqq&\; \epsilon^{abcd}\epsilon^{a'}{}_{b'cd}\phi_a\phi_{a'}\overset{(1)}{\Phi}{}^{b'}{}_{b} \notag \\
	=&\; \epsilon^{abcd}\epsilon^{a'b'}{}_{cd}\phi_a\phi_{a'}\nabla_b \phi_{b'}, \\
	\mathcal{L}_4^{\rm gal\phi}\coloneqq&\;  \epsilon^{abcd}\epsilon^{a'}{}_{b'c'd}\phi_a\phi_{a'}\overset{(1)}{\Phi}{}^{b'}{}_{b}\overset{(1)}{\Phi}{}^{c'}{}_{c}\notag \\
	=&\;  \epsilon^{abcd}\epsilon^{a'b'c'}{}_{d}\phi_a\phi_{a'}\nabla_b \phi_{b'}\nabla_c \phi_{c'}.\label{eq:shim-AokiG4}
\end{align}
Note in passing that the second term can also be written using the epsilon tensor [see Eqs.~\eqref{et01} and~\eqref{et02}].
Integrating out the connection, one can rewrite the above theory into an equivalent one in the metric formalism. It can then be shown that
the resultant Lagrangian belongs to the DHOST family~\cite{Langlois:2015cwa,Crisostomi:2016czh,BenAchour:2016cay,BenAchour:2016fzp,Takahashi:2017pje,Langlois:2018jdg,Takahashi:2021ttd} (see also Appendix~\ref{app:DHOST}). Thus, one concludes that the metric-affine theory with the projective-invariant Lagrangian~\eqref{Lag:shim-aoki} is free from ghost degrees of freedom.

Two comments are now in order. First, the Lagrangian~\eqref{Lag:shim-aoki} does not in general reduce to a Horndeski Lagrangian in the $\kappa^a{}_{bc}\to 0$ limit; we need to tune the coefficients of the various terms in Eq.~\eqref{Lag:shim-aoki} to reproduce the Horndeski theory. In this sense, Eq.~\eqref{Lag:shim-aoki} gives a metric-affine extension of something beyond Horndeski. Nevertheless, after integrating out the connection to rewrite the theory in the metric formalism, we arrive at a ghost-free theory in the DHOST family anyway.
In other words, our starting point is not necessarily a ghost-free metric theory.
Second, one could consider the Lagrangian where $\overset{(1)}{\Phi}{}^{c'}{}_{c}$ in Eq.~\eqref{eq:shim-AokiG4} is replaced with $\overset{(1)}{\Phi}{}^{c}{}_{c'}$ as
\begin{align}
    \mathcal{L}_4^{\rm gal\phi'}\coloneqq&\;  \epsilon^{abcd}\epsilon^{a'}{}_{b'c'd}\phi_a\phi_{a'}\overset{(1)}{\Phi}{}^{b'}{}_{b}\overset{(1)}{\Phi}{}^{c}{}_{c'} \notag \\
	=&\;  \epsilon^{abcd}\epsilon^{a'b'c'}{}_{d}\phi_a\phi_{a'}\nabla_b \phi_{b'}\nabla_{c'} \phi_{c},
\end{align}
instead of ${\mathcal{L}_4^{\rm gal\phi}}$ in the Lagrangian~\eqref{Lag:shim-aoki}, but one ends up with the same Lagrangian in the metric formalism whichever Lagrangian is used as a starting point in the metric-affine formalism~\cite{Aoki:2018lwx}.
This essentially comes from the fact that $\phi_a$ is not a generic vector but is a gradient of the scalar field.
When we discuss vector-tensor theories in the metric-affine formalism in Sec.~\ref{sec:epsilonVT}, this point needs special care.

So far we have used only a single building block, $\overset{(1)}{\Phi}{}^a{}_b$, among the four terms listed in Eq.~\eqref{eq:TermsOfPhi}.
However, it is easy to build a projective-invariant term by invoking another term from the remaining three, which leads to our second method toward a projective-invariant Lagrangian.
Suppose that we use $\overset{(1)}{\Phi}{}^a{}_b$ and $\overset{(2)}{\Phi}{}^a{}_b$.
We then immediately find a projective-invariant second derivative, $\overset{(1)}{\Phi}{}^a{}_b+\overset{(2)}{\Phi}{}^a{}_b$, and so its trace,
\begin{align}
\overset{(1)}{\Phi}{}^a{}_a+\overset{(2)}{\Phi}{}^a{}_a=
	\nabla^a\phi_a+\nabla_a\phi^a,
\end{align}
is a projective-invariant combination that reduces to $D^a\phi_a$ in the $\kappa^a{}_{bc}\to 0$ limit.
Indeed, one may instead consider different projective-invariant second derivatives, $\overset{(1)}{\Phi}{}^a{}_b+\overset{(3)}{\Phi}{}^a{}_b$ and $\overset{(1)}{\Phi}{}^a{}_b+\overset{(4)}{\Phi}{}^a{}_b$, but, for simplicity, we would like to concentrate on the case of using only the combination $\overset{(1)}{\Phi}{}^a{}_b+\overset{(2)}{\Phi}{}^a{}_b$ in this paper.

For instance, such building blocks have been adopted in Ref.~\cite{Helpin:2019vrv} to study the instability of metric-affine theory with the Lagrangian
\begin{align}
    L=\sum^4_{i=2}\mathcal{L}_{i}^{\mathrm{MA}},
\end{align}
with
\begin{align}
    \mathcal{L}_2^{\mathrm{MA}} &\coloneqq G_2,\\
    \mathcal{L}_3^{\mathrm{MA}} &\coloneqq G_3^{(1)} \overset{(1)}{\Phi}+G_3^{(2)}\overset{(2)}{\Phi}+G_3^{(3)}\overset{(3)}{\Phi}+G_3^{(4)}\overset{(1)}{\Phi}{}^{ab}\phi_a\phi_b+G_3^{(5)}\overset{(2)}{\Phi}{}^{ab}\phi_a\phi_b,\\
    \mathcal{L}_4^{\mathrm{MA}} &\coloneqq G_4(\phi)R,
\end{align}
where $\overset{(i)}{\Phi}\coloneqq \overset{(i)}{\Phi}{}^a{}_a$ ($i=1,2,3$), $G_2$ and $G_3^{(i)}$ ($i=1,2,3,4,5$) are arbitrary functions of $\phi$ and $\mathcal{X}$, and $G_4$ is arbitrary function of $\phi$. Note that, in this case, we need the additional constraint that $\mathcal{L}_3^{\mathrm{MA}}$ is projective invariant since $\mathcal{L}_3^{\mathrm{MA}}$ is not projective invariant alone.
After integrating the connection out from this Lagrangian, one can see that this Lagrangian belongs to the Horndeski family in the metric formalism~\cite{Helpin:2019vrv}.

Of course, there are many other possible ways to extend the Horndeski theory in the metric formalism to the metric-affine formalism, but we will not discuss them in this paper.

In summary, these examples of ghost-free theories are the result of imposing either the torsionless condition or the projective symmetry.
These studies suggest that in the metric-affine formalism, it is useful to impose either the torsionless condition (which is especially called the Palatini formalism) or the projective invariance to avoid ghost occurrence.

We will spend the rest of this paper discussing vector-tensor theories in the metric-affine formalism.
Then, with the help of the results on scalar-tensor theories in the metric-affine formalism, as seen in this section, we expect that the conditions between the coefficient functions needed to obtain ghost-free theories will be relaxed in the metric-affine formalism.
Therefore, the following calculations will be based on a slight extension of the generalized Proca theory with independent coefficient functions.

\section{Vector-tensor theories in the Palatini formalism}\label{sec:VTInPalatini}

Let us now start to explore vector-tensor theories within the Palatini formalism.
In this section, we assume that the torsion vanishes at the level of the Lagrangian,
\begin{align}
	T^a{}_{bc}=0,
\end{align}
and integral out the connection to see what the resultant theory is in the metric formalism.

We start by considering the following general Lagrangian,
\begin{align}
	L&=G_2(X)+G_3(X)g^{ab}\nabla_aA_b+G_4(X)R\notag\\
	&+H_1(X)(\nabla^a A_a)^2+H_2(X)g^{ad}g^{bc}\nabla_a A_b\nabla_c A_d+H_3(X)g^{ac}g^{bd}\nabla_a A_b\nabla_c A_d \label{eq:LagrangianInPalatini},
\end{align}
where $X\coloneqq-A^aA_a/2$. This Lagrangian contains up to quadratic terms in the connection, and hence the Euler-Lagrange equations for the connection, $\mathcal{P}_a{}^{bc}=0$, depend linearly on the connection.
Therefore, the equation $\mathcal{P}_a{}^{bc}=0$ can be solved to give
\begin{align}
    \kappa_{abc}&=C_0A_aA_bA_c+C_1g_{bc}A_a+2C_2g_{a(b}A_{c)}+2C_3A_aD_{(b}A_{c)}+2C_4A_{(b}D_{c)}A_a+2C_5A_{(b}D_{|a|}A_{c)}\notag\\
	&\quad+2C_6A^dA_aA_{(b}D_{c)}A_d+C_{7}A^dA_bA_cD_{a}A_d+2C_{8}A_d A_aA_{(b}D^{d}A_{c)}+C_{9}A^dA_bA_cD_{d}A_a\notag\\
	&\quad+2C_{10}A^d g_{a(b}D_{c)}A_d+C_{11}A^d g_{bc}D_aA_d+2C_{12}A_d g_{a(b}D^{d}A_{c)}+C_{13}A^d g_{bc}D_dA_a,\label{kappa-soln}
\end{align}
where the explicit forms of the coefficients $C_0,\cdots, C_{13}$ are presented in Appendix~\ref{app:ExplicitFormPalatini}.
To obtain the solution~\eqref{kappa-soln} we have assumed that
\begin{align}
    G_4\left(G_4-2H_{23}X\right)
    \left(3G_4+2H_{23}X\right)
    \left[
    3G_4^2+4G_4\left(-2H_1+H_{23}\right)X-8H_{23}\left(4H_1+H_{23}\right)X^2
    \right]\neq 0,
\end{align}
with $H_{23}:=H_2+H_3$.
Substituting the above solution to Eq.~\eqref{eq:LagrangianInPalatini}, we obtain the Lagrangian for a vector-tensor theory in the metric formalism as
\begin{align}
     L&=g_2(X)+g_3(X)D^aA_a+h_3(X)A^aA^bD_aA_b+g_4(X)\overset{g}{R}+\sum^9_{i=1}\alpha_i(X)L_i \label{eq:IntegrateOutInVT},
\end{align}
with
\begin{align}
    L_1&=g^{a(c}g^{d)b}D_aA_bD_cA_d,
    \quad
    L_2=g^{ab}g^{cd}D_aA_bD_cA_d,\notag \\
    L_3&=\left(A^{a} A^{b} g^{c d}+A^{c} A^{d} g^{a b}\right)D_aA_bD_cA_d,
    \quad
    L_4=\left(A^{a} A^{(c} g^{d) b}+A^{b} A^{(c} g^{d) a}\right)D_aA_bD_cA_d,
    \notag \\
    L_5&=A^{a} A^{b} A^{c} A^{d}D_aA_bD_cA_d,
    \quad
    L_6=g^{a[c} g^{d]b}D_aA_bD_cA_d,
    \quad
    L_7=\left(A^{a} A^{[c} g^{d] b}-A^{b} A^{[c} g^{d] a}\right)D_aA_bD_cA_d,\notag \\
    L_8&=\left(A^{a} A^{c} g^{b d}-A^{b} A^{d} g^{a c}\right)D_aA_bD_cA_d,
    \quad
    L_9=\epsilon^{a b c d}D_aA_bD_cA_d,
\end{align}
where $h_3$, $g_i$ ($i=2,3,4$), and $\alpha_i$ ($i=1, \dots, 9$) are written in terms of $G_i(X)$ and $H_i(X)$ and hence are the functions of $X$.
The explicit forms of these coefficients are given in Appendix~\ref{app:ExplicitFormPalatini}.

Note that $D_aA_b\neq D_bA_a$ in vector-tensor theories, unlike the case in scalar-tensor theories where $D_a\phi_b=D_b\phi_a$.
This means that even though the solution for the connection depends on $H_2$ and $H_3$ through the combination $H_{23}$ and hence is symmetric under the exchange of $H_2$ and $H_3$, the vector-tensor theory in the metric formalism obtained after integrating out the connection is not so.
This is easily seen from the fact that the Lagrangian contains
\begin{align}
    L\supset \alpha_6L_6,
    \quad \alpha_6=H_3-H_2,
\end{align}
(see also Appendix~\ref{app:ExplicitFormPalatini}).

Having thus obtained the Lagrangian for a vector-tensor theory in the metric formalism, we now discuss whether or not the resultant theory is free from the Ostrogradsky ghost by checking the degeneracy conditions.
The degeneracy conditions read
\begin{align}
    D_0=D_1=D_2=0 \label{eq:DegenerateConditionsForEVT},
\end{align}
(see Appendix~\ref{sec:EVT} for the concrete expression for $D_0$, $D_1$, and $D_2$).
These conditions impose some relations among the functions $g_4$ and $\alpha_i$, which then translate to the relations among $G_4$ and $H_i$. More explicitly, the condition $D_0=0$ reduces to
\begin{align}
	&\left(G_4-X G_{4 X}\right) \left[H_3 \left(3
	G_4+4 H_3 X\right)-4 H_2^2 X\right]\times\notag\\
	&\left[-G_4 \left(10 H_1-H_{23}\right) H_{23} X+3 G_4^2 \left(H_1+H_{23}\right)-6
    H_{23}^2 \left(4 H_1+H_{23}\right) X^2\right]=0 \label{eq:D0=0InPalatini}.
\end{align}

Let us first consider the branch $G_4-X G_{4 X}=0$, in which the condition is solved to give $G_4=CX$ with $C$ being a nonvanishing constant. Then, one can see that $D_1=0$ is automatically satisfied in this branch.
The condition $D_2=0$ reduces to
\begin{align}
    \left[H_3 \left(3 C+H_3\right)-H_2^2\right]
	\left[3 C^2+C \left(-17 H_1+H_{23}\right)-9
	H_{23} \left(4
	H_1+H_{23}\right)\right]=0.\label{b1-D20}
\end{align}
Therefore, the degeneracy condition $D_2=0$ imposes a certain constraint among $H_1$, $H_2$, and $H_3$.
Then, even if some of the three functions are arbitrary functions, we can get healthy theories.

Here, except for the special cases where $C$ is zero, there are obviously two different solutions.
First, in the branch of $H_3 \left(3 C+H_3\right)-H_2^2=0$, $H_2$ is solved as
\begin{align}
	H_2=\pm \sqrt{H_3}\sqrt{3C+H_3},
\end{align}
where we assume $H_3(3C+H_3)>0$.
Then, since the degeneracy conditions are satisfied for any forms of $H_1$ and $H_3$, one can obtain healthy theories without dependency on the forms of $H_1$ and $H_3$.

Next, in the other branch of $D_2=0$, $H_1$ can be uniquely solved as
\begin{align}
	H_1=\frac{3 C^2+C H_{23}-9
	H_{23}}{17 C+36 H_{23}},
\end{align}
except for the $17 C+36 H_{23}=0$ case.
Here the degeneracy conditions are satisfied for any forms of $H_2$ and $H_3$.

Let us come back to the degeneracy condition $D_0=0$ [i.e., Eq.~\eqref{eq:D0=0InPalatini}], and consider another branch where 
$H_3(3G_4+4H_3X)-4H_2^2X=0$ is satisfied.
In the special case with $H_3=0$, it is immediately found that $H_2=0$.
In this case, all the degeneracy conditions turn out to be satisfied, then $G_4$ and $H_1$ are allowed to be arbitrary functions.
In the generic case with $H_3\neq 0$, we have 
\begin{align}
    G_4=\frac{4XH_{23}(H_2-H_3)}{3H_3}\,(\neq 0)\label{eq:PaltiniG4}.
\end{align}
Substituting this to the second degeneracy condition $D_1=0$, we obtain 
\begin{align}
    \left[6 H_2^2+11 H_3 H_2+5 H_3^2+H_1 \left(6 H_2+13
	H_3\right)\right] 
	\left[\frac{H_{23}(H_2-H_3)}{H_3}
	\right]_{,X}=0.
\end{align}
Here, recalling Eq.~\eqref{eq:PaltiniG4}, we can obtain $G_4=CX$ if $[H_{23}(H_2-H_3)/H_3]_{,X}=0$.
Hence $H_{23}(H_2-H_3)/H_3=3C/4$.
With this branch, the rest degeneracy condition $D_2=0$ is automatically satisfied.
In this situation, the theory is ghost-free for any forms of $H_2$ and $H_3$.
Next, let us concentrate on the branch which satisfies
\begin{align}
	6 H_2^2+11 H_3 H_2+5 H_3^2+H_1 \left(6 H_2+13H_3\right)=0,
\end{align}
this can be uniquely solved for $H_1$ as
\begin{align}
	H_1=-\frac{6 H_2^2+11 H_3 H_2+5 H_3^2}{6 H_2+13H_3},
\end{align}
except for the $6 H_2+13H_3=0$ case.
However, the form of degeneracy condition $D_2=0$ of this branch is messy.
Therefore, for simplicity, we will assume the following ansatz for $H_2$ and $H_3$:
\begin{align}
	H_2=aX^n, \quad H_3=bX^n,
\end{align}
where $a$, $b$ and $n$ is constant.
By this ansatz, we can find the ratio of $a$ to $b$ for each value of $n$ from the condition that $D_2=0$. For example, when $n=1$, the real solution is
\begin{align}
    a/b\simeq -4.46,\quad -3.90,\quad -0.300,\quad -0.0463,
\end{align}
and when $n=2$, the real solution is
\begin{align}
   a/b\simeq -4.48,\quad -3.76,\quad -0.311,\quad -0.0462.
\end{align}
In this way, it is possible to obtain a numerical solution by specifying the value of $n$.
In any case, we can obtain ghost-free theories without dependency on the forms of some of the coefficient functions.

Let us finally come back to the degeneracy condition $D_0=0$ [i.e., Eq.~\eqref{eq:D0=0InPalatini}] and focus on the third branch where
\begin{align}
    -G_4 \left(10 H_1-H_{23}\right) H_{23} X+3 G_4^2 \left(H_1+H_{23}\right)-6
   H_{23}^2 \left(4 H_1+H_{23}\right) X^2=0.
\end{align}
Then, we can solve it for $H_1$ as
\begin{align}
    H_1=\frac{H_{23} \left(G_4 H_{23} X+3 G_4^2-6 H_{23}^2 X^2\right)}{10 G_4 H_{23} X-3 G_4^2+24 H_{23}^2 X^2},
\end{align}
except for the $10 G_4 H_{23} X-3 G_4^2+24 H_{23}^2 X^2=0$ case.
One can see that $D_1=0$ reduces to
\begin{align}
    &H_{23} \left(G_4-X G_{4 X}\right) \times\notag\\
    &\left(-4 G_4 \left(H_2-22 H_3\right)
   H_{23}^2 X^2+G_4^2 H_{23} \left(5 H_2+103 H_3\right) X+33 G_4^3 H_3-12
   \left(H_2-H_3\right) H_{23}^3 X^3\right)=0,
\end{align}
and hence $D_2=0$ reduces to
\begin{align}
    &\left(H_3 \left(3 G_4+H_3 X\right)-H_2^2 X\right)\times\notag\\
    &\Big[-G_4^2 H_{23}^2 X^2 \left(8
   H_{23} G_{4 X}+13 G_{4 X}^2+36 H_{23}^2\right)-4 G_4 H_{23}^3
   X^3 G_{4 X} \left(5 G_{4 X}-18 H_{23}\right)-36 H_{23}^4 X^4 G_{4 X}^2\notag\\
   &\quad+G_4^4
   \left(18 H_{23} G_{4 X}+9 G_{4 X}^2+31 H_{23}^2\right)+2 G_4^3
   H_{23} X \left(-H_{23} G_{4 X}+3 G_{4 X}^2+14
   H_{23}^2\right)\Big]=0.
\end{align}
It is clear that the above degeneracy conditions give some constraints among $G_4$, $H_2$, and $H_3$, then some of the three functions are arbitrary.

To sum up, with the above analysis, we find that the degeneracy of the system gives some constraints on the four arbitrary functions (i.e., $G_4$, $H_1$, $H_2$, and $H_3$).
We then can obtain ghost-free theories without dependency on the forms of some of the ones.

\section{Vector-tensor theories in the metric-affine formalism}\label{sec:VTInMA}

In this section, we extend the generalized Proca theory in the metric formalism to the metric-affine formalism by imposing the projective invariance.

Following Eq.~\eqref{eq:TermsOfPhi}, for a first derivative of a vector field $A_\mu$ we may define four independent quantities built out of $\nabla_\mu$ and the metric,
\begin{align}
    \overset{(1)}{\mathcal{A}}{}^\mu{}_\nu=\nabla^\mu A_\nu, \quad
        \overset{(2)}{\mathcal{A}}{}^\mu{}_\nu=g_{\nu\rho}\nabla^\mu A^\rho, \quad 
    \overset{(3)}{\mathcal{A}}{}^\mu{}_\nu=(1/4)g_{\alpha\beta}\nabla^\mu(g^{\alpha\beta}A_\nu), \quad
        \overset{(4)}{\mathcal{A}}{}^\mu{}_\nu=\nabla_\rho(g^{\mu\rho}A_\nu),
\end{align}
all of which reduce to $D^\mu A_\nu$ in the $\kappa^a{}_{bc}\to 0$ limit.
Under a projective transformation, these quantities transform in the same way as Eq.~\eqref{phi-p-tr}:
\begin{align}
    \overset{(1)}{\mathcal{A}}{}^\mu{}_\nu\to
    \overset{(1)}{\mathcal{A}}{}^\mu{}_\nu-\xi^\mu A_\nu,
    \quad
    \overset{(2)}{\mathcal{A}}{}^\mu{}_\nu\to
    \overset{(2)}{\mathcal{A}}{}^\mu{}_\nu+\xi^\mu A_\nu,
    \quad
    \overset{(3)}{\mathcal{A}}{}^\mu{}_\nu\to
    \overset{(3)}{\mathcal{A}}{}^\mu{}_\nu+\xi^\mu A_\nu,
    \quad
    \overset{(4)}{\mathcal{A}}{}^\mu{}_\nu\to
    \overset{(4)}{\mathcal{A}}{}^\mu{}_\nu+\xi^\mu A_\nu.
\end{align}

As reviewed in Sec.~\ref{sec:STInMA}, in the case of scalar-tensor theories, we can systematically construct a projective-invariant Lagrangian by contracting the epsilon tensors with $\phi_a$ and $\overset{(1)}{\Phi}{}^a{}_b$, which always yields, after integrating out the connection, a DHOST theory in the metric formalism.
One can also use two of the four different second derivatives (say, $\overset{(1)}{\Phi}{}^a{}_b$ and $\overset{(2)}{\Phi}{}^a{}_b$) to build a projective-invariant Lagrangian.
Here we will follow the same approaches in the case of vector-tensor theories as well and see how healthy theories in the metric formalism are obtained after integrating out the connection.

\subsection{Contracting the epsilon tensor}\label{sec:epsilonVT}

We start from the Lagrangian
\begin{align}
	L=G_4(X)g^{ab}R_{ab}+F(X)G^{ab}A_aA_b+G_2(X)+G_3(X) \mathcal{L}_3^{\rm gal}+H_1(X) \mathcal{L}_4^{\rm gal}+H_2(X) \mathcal{L}_4^{\rm gal'},
	\label{lag:epsironLA}
\end{align}
where we introduced the terms obtained by contracting the epsilon tensors with the vector field and its derivatives as
\begin{align}
	\mathcal{L}_3^{\rm gal} \coloneqq&\;  \epsilon^{abcd}\epsilon^{a'}{}_{b'cd}A_aA_{a'}\overset{(1)}{\mathcal{A}}{}^{b'}{}_{b}\\
	=&\; \epsilon^{abcd}\epsilon^{a'b'}{}_{cd}A_aA_{a'}\nabla_b A_{b'}, \\
	\mathcal{L}_4^{\rm gal}\coloneqq&\;  \epsilon^{abcd}\epsilon^{a'}{}_{b'c'd}A_aA_{a'}\overset{(1)}{\mathcal{A}}{}^{b'}{}_{b}\overset{(1)}{\mathcal{A}}{}^{c'}{}_{c} \\
	=&\;  \epsilon^{abcd}\epsilon^{a'b'c'}{}_{d}A_aA_{a'}\nabla_b A_{b'}\nabla_c A_{c'}, \\
	\mathcal{L}_4^{\rm gal'}\coloneqq&\;  \epsilon^{abcd}\epsilon^{a'}{}_{b'c'd}A_aA_{a'}\overset{(1)}{\mathcal{A}}{}^{b'}{}_{b}\overset{(1)}{\mathcal{A}}{}^{c}{}_{c'}\\
	=&\;  \epsilon^{abcd}\epsilon^{a'b'c'}{}_{d}A_aA_{a'}\nabla_b A_{b'}\nabla_{c'} A_{c}.
\end{align}
It is easy to see that each term in the Lagrangian is projective invariant.
The Lagrangian~\eqref{lag:epsironLA} is thought of as a vector-tensor generalization of a scalar-tensor theory in the metric-affine formalism developed in Ref.~\cite{Aoki:2018lwx}.
The vector nature as compared to the scalar case is manifest in the inclusion of the last two terms.

Since the above Lagrangian is a quadratic polynomial with respect to the connection, the Euler-Lagrange equation for $\kappa^a{}_{bc}$, ${\cal P}_a{}^{bc}=0$, is a linear algebraic equation.
One can write the generic form of $\kappa^a{}_{bc}$ as
\begin{align}
    \kappa_{abc}&=
    C_0A_aA_bA_c+
    C_1A_ag_{bc}+C_2A_cg_{ab}+C_3A_aD_bA_c+C_4A_bD_cA_a+C_5A_cD_aA_b
    \notag \\ &\quad 
    +C_6A_aD_cA_b+C_7A_bD_aA_c
    +C_8A_cD_bA_a
    +C_9A_aA_bA^dD_cA_d+C_{10}A_bA_cA^dD_aA_d+C_{11}A_cA_aA^dD_bA_d
    \notag \\ &\quad 
    +C_{12}A_aA_bA^dD_dA_c
    +C_{13}A_bA_cA^dD_dA_a
    +C_{14}A_cA_aA^dD_dA_b
    \notag \\ &\quad 
    +C_{15}g_{ab}A^dD_cA_d+C_{16}g_{bc}A^dD_aA_d
    +C_{17}g_{ab}A^dD_dA_c+C_{18}g_{bc}A^dD_dA_a,\label{gen:kappa}
\end{align}
where, thanks to the projective invariance, the coefficient functions of terms proportional to $g_{ac}$ are automatically equal to zero.
If one does not impose the projective invariance on the theory, one must take such terms into account.
Substituting this to ${\cal P}_a{}^{bc}=0$, one can determine the coefficients $C_j$ ($j=0,\dots,18$) and express them in terms of $X$, $D^aA_a$, $A^aA^bD_aA_b$, $F(X)$, $G_3(X)$, $G_4(X)$, $H_1(X)$, and $H_2(X)$.
The explicit forms of these coefficients are presented in Appendix~\ref{app:ExplicitFormEpsilon}.
Here, to obtain the explicit forms of $C_j$, we have assumed that the combination of denominators of terms appearing in Appendix~\ref{app:ExplicitFormEpsilon} are nonzero as in the case of the previous section (e.g., $G_4\neq0$ is assumed in the following).
Now, using the solution for $\kappa^a{}_{bc}$, one can remove the connection from the Lagrangian~\eqref{lag:epsironLA} and obtain the reduced Lagrangian in the form of Eq.~\eqref{eq:IntegrateOutInVT} in the metric formalism.
One can then check explicitly that the resultant theory automatically satisfies the vector-tensor degeneracy conditions~\eqref{eq:DegenerateConditionsForEVT}.
Thus, as in the case of scalar-tensor theories, a projective-invariant vector-tensor Lagrangian in the metric-affine formalism obtained systematically by contracting the epsilon tensor always yields, after integrating out the connection, a degenerate vector-tensor theory in the metric formalism, and hence is healthy.

\subsection{Projective-invariant Lagrangian obtained in a different way}\label{sec:minimalVT}

Next, let us start from the Lagrangian
\begin{align}
	L=G_4(X)g^{ab}R_{ab}+G_2(X)+G_3 (X)\mathcal{L}_3+\sum_{i=1}^3 Q_i(X)\mathcal{L}_4^i,
\end{align}
where
\begin{align}
	\mathcal{L}_3&\coloneqq \mathcal{A}{}^{a}{}_{a}
	=(\nabla_aA^a+\nabla^aA_a)/2, \\
	\mathcal{L}_4^1&\coloneqq \delta_a^b\delta_c^d \mathcal{A}{}^{a}{}_{b}\mathcal{A}{}^{c}{}_{d}
	=\frac{1}{4}[(\nabla_a A^a)^2+2(\nabla_a A^a)(\nabla^b A_b)+(\nabla^a A_{a})^2], \\
	\mathcal{L}_4^2 &\coloneqq \delta_a^b\delta_c^d \mathcal{A}{}^{a}{}_{d}\mathcal{A}{}^{c}{}_{b}
	=\frac{1}{4}[g^{ad}g^{bc}\nabla_{a}A_{b}\nabla_{c}A_{d}+2\nabla_{a}A_{b}\nabla^{b}A^{a}+g_{ad}g_{bc}\nabla^a A^b\nabla^c A^d], \\
	\mathcal{L}_4^3 &\coloneqq g_{ac}g^{bd} \mathcal{A}{}^{a}{}_{b}\mathcal{A}{}^{c}{}_{d}=\frac{1}{4}[g^{ac}g^{bd}\nabla_{a}A_{b}\nabla_{c}A_{d}+2\nabla_{a}A_{b}\nabla^{a}A^{b}+g_{ac}g_{bd}\nabla^a A^b\nabla^c A^d],
\end{align}
with $\mathcal{A}{}^{a}{}_{b}\coloneqq (\overset{(1)}{\mathcal{A}}{}^a{}_b+\overset{(2)}{\mathcal{A}}{}^a{}_b)/2$ being a projective-invariant combination. The Lagrangian depends on the first derivatives of the vector field through this particular projective-invariant combination.

We again write the generic form of $\kappa^a{}_{bc}$ as Eq.~\eqref{gen:kappa}, and use the Euler-Lagrange equation for the connection to determine the coefficients $C_j$. The explicit expressions for $C_j$, in this case, are shown in Appendix~\ref{app:ExplicitFormMinimal}.
To obtain $C_j$, we assumed that the combination of denominators of terms appearing in Appendix~\ref{app:ExplicitFormMinimal} are nonzero [e.g., $G_4(G_4-Q_3 X)\neq0$ in the following].
Integrating out the connection, we arrive at a vector-tensor Lagrangian in the metric formalism in the form of Eq.~\eqref{eq:IntegrateOutInVT}.
In the present case, however, the degeneracy conditions are not satisfied automatically. One of the degeneracy conditions, $D_0=0$, reads
\begin{align}
	Q_3=0.
\end{align}
If $Q_3=0$, the other degeneracy conditions, $D_1=D_2=0$, are satisfied automatically.
The physically interesting family of degenerate theories is thus obtained provided that $Q_3=0$.
In particular, we have a Case A extended vector-tensor theory in the terminology of Ref.~\cite{Kimura:2016rzw} if, in addition, $Q_2=-Q_1$ is satisfied.

\section{Conclusions}\label{sec:concl}

In this paper, we have considered metric-affine extensions of vector-tensor theories and studied their viability by inspecting the vector-tensor theories in the metric formalism obtained after integrating out the connection.
For this to be doable we have focused on the case where the Lagrangian is at most quadratic in the first derivatives of the vector field.

Previous studies imply that imposing the torsionless condition $T^a{}_{bc}=0$ from the beginning (dubbed as the Palatini formalism) or imposing the invariance under the projective transformation $\Gamma^a{}_{bc}\to\Gamma^a{}_{bc}+\delta^a_c\xi_b$ is helpful for obtaining ghost-free metric-affine theories.
In this paper, we have followed the same idea and considered (i) vector-tensor theories in the Palatini formalism ($T^a{}_{bc}=0$); (ii) projective-invariant vector-tensor theories in the metric-affine formalism constructed systematically by contracting various terms with the completely antisymmetric tensor $\epsilon^{abcd}$; and (iii) projective-invariant vector-tensor theories in the metric-affine formalism constructed in an ad-hoc manner by using a certain projective-invariant first derivative of the vector field as a building block.
The cases (i) and (ii) are extensions of the works~\cite{Helpin:2019kcq} and~\cite{Aoki:2018lwx}, respectively, to vector-tensor theories.
On the other hand, the case (iii) is an approach completely independent of them.

In all the above three cases, we have integrated out the connection to write the corresponding Lagrangians in the metric formalism, and then investigated the degeneracy conditions~\cite{Kimura:2016rzw} for the resultant vector-tensor Lagrangians.
As the first of our work, we have concentrated on the case called ``the Palatini formalism,'' where the torsionless condition is imposed at the beginning.
Under the settings, we have considered a general vector-tensor theory (including such as generalized Proca theory) in the Palatini formalism.
Then we have investigated the conditions that the theory is degenerate.
In this case, we have seen that the degeneracy conditions are messy.
For some of the branches, we have indeed solved the degeneracy conditions analytically or numerically.
To sum up the results, one can see that some of the four arbitrary functions in the Lagrangian (except for $G_2$ or $G_3$ terms) are determined by a degeneracy.
Also, we have studied vector-tensor theories in the metric-affine formalism.
We then have required that the theories are projective invariant.
To construct a projective-invariant Lagrangian, we have considered two approaches.
First, we have constructed the Lagrangian by use of the epsilon tensor, which corresponds to a vector-tensor version of Ref.~\cite{Aoki:2018lwx}.
In this case, the degeneracy conditions are immediately satisfied.
Second, we have obtained a projective-invariant Lagrangian in a different way, where at first projective-invariant combinations are used.
In this case, we have arrived that degeneracy conditions are satisfied as long as $Q_3=0$.

We thus conclude that, while the torsionless condition and the projective invariance do not guarantee the ghost-free nature of the theories, the systematic construction of projective-invariant theories is a promising way to arrive at healthy metric-affine theories since in metric-affine geometry, imposing the projective invariance would be a more prospective manner to get ghost-free (vector-tensor) theories than imposing the torsionless condition.
We note that we have investigated only up to quadratic terms of a connection, so it would be interesting to add, e.g., higher curvature terms to action.
We hope to come back to these issues in the near future.

\acknowledgments
We are grateful to Tsutomu Kobayashi for fruitful discussions and comments at the early stage of the work.
We would like to thank Katsuki Aoki for the helpful discussions.
The work of TI was supported by the Rikkyo University Special Fund for Research.

\appendix

\section{Degenerate scalar-tensor theories in the metric formalism}
In this section, we only discuss theories in the metric formalism, so all quantities related to curvature are defined by the Levi-Civita connection, but omit the $g$ sign for simplification of notation.

\subsection{Horndeski theory}\label{app:Horndeski}

The Horndeski Lagrangian is given by~\cite{Horndeski:1974wa,Deffayet:2011gz,Kobayashi:2011nu}
\begin{align}
    L=\sum_{i=2}^5\mathcal{L}_i,
\end{align}
where
\begin{align}
	\mathcal{L}_{2} &\coloneqq G_{2}(\phi, \mathcal{X}), \\
	\mathcal{L}_{3} &\coloneqq G_{3}(\phi, \mathcal{X}) D^{a} \phi_{a}, \\
	\mathcal{L}_{4} &\coloneqq G_{4}(\phi, \mathcal{X}) R+G_{4\mathcal{X}}\left[(D^{a} \phi_{a})^{2}-D_{a} \phi_{b} D^{b} \phi^{a}\right], \\
\mathcal{L}_{5} &\coloneqq G_{5}(\phi, \mathcal{X}) G_{a b} D^{a} \phi^{b}-\frac{1}{6} G_{5\mathcal{X}}\left[(D^a \phi_a)^{3}+2 D_{a} \phi_{b} D^{c} \phi^{a} D^{b} \phi_{c}-3(D^c \phi_c) D_{a} \phi_{b} D^{a} \phi^{b}\right].
\end{align}
Here, $\mathcal{X}\coloneqq -g^{ab}\phi_a\phi_b/2$ with $\phi_a \coloneqq D_a \phi$.
The Euler-Lagrange equations are second-order differential equations, so they are free from the Ostrogradsky instability.

One can rewrite the Horndeski Lagrangian in the language of the epsilon tensor.
As is clear from Eq.~\eqref{eq:HorndeskiL3epsilonChange}, we can do for $\mathcal{L}_3$.
Therefore, we would like to take $\mathcal{L}_4$ as an example.
Since the contribution of the $\phi$ dependence of $G_4$ can be absorbed by the redefinitions of $G_2$ and $G_3$, we now for simplicity consider the case where the shift symmetry is imposed and $G_4$ depends on only $\mathcal{X}$.
Then, one can rewrite $\mathcal{L}_4$ as
\begin{align}
    \mathcal{L}_4&=G_4 R+G_{4\mathcal{X}}\left[(D^{a} \phi_{a})^{2}-D_{a} \phi_{b} D^{b} \phi^{a}\right]\\
    &=H_4 G^{ab}\phi_a \phi_b+\frac{1}{2}H_{4\mathcal{X}}\epsilon^{abcd}\epsilon^{a'b'c'}{}_d\phi_a\phi_{a'}\phi_{bb'}\phi_{cc'}+(\text{total derivative}),
\end{align}
with
\begin{align}
    G_4=H_4\mathcal{X}.
\end{align}

Note that the coefficient function of the counterterm, $G_{4X}$, was rewritten as well.

\subsection{Degenerate scalar-tensor theories}\label{app:DHOST}

In this section, we focus on the subclass of DHOST theories whose Lagrangian is up to a quadratic polynomial in $D_a\phi_b$, which are called quadratic DHOST theories.
The subclass of DHOST theories is given by
\begin{align}
    L=f(\phi,\mathcal{X})R+G_2(\phi,\mathcal{X})+G_3(\phi,\mathcal{X})\square \phi+C^{abcd}D_a \phi_b D_c \phi_d,
\end{align}
where
\begin{align}
    C^{abcd} \coloneqq &\; \frac{1}{2} \alpha_{1}\left(g^{a c} g^{b d}+g^{a d} g^{b c}\right)+\alpha_{2} g^{a b} g^{c d}+\frac{1}{2} \alpha_{3}\left(\phi^a \phi^b g^{c d}+\phi^c \phi^d g^{a b}\right) \notag\\
    &+\frac{1}{4} \alpha_{4}\left(\phi^a \phi^c g^{b d}+ \phi^b \phi^c g^{a d}+\phi^a \phi^d g^{b c}+\phi^b \phi^d g^{a c}\right) +\alpha_{5} \phi^a \phi^b \phi^c \phi^d,
\end{align}
and, the Horndeski theory is a special case with $\alpha_1=-\alpha_2=-f_\mathcal{X}$ and $\alpha_3=\alpha_4=\alpha_5=0$.

Note that this Lagrangian has the higher-order Euler-Lagrange equations, which lead to the Ostragaldsky instability.
To avoid this, we should impose the degeneracy conditions, which require that the determinant of the Hessian vanishes.
The degeneracy conditions are summarized as
\begin{align}
    D_0(\phi,\mathcal{X})=D_1(\phi,\mathcal{X})=D_2(\phi,\mathcal{X})=0,
\end{align}
where the details of $D_i$ ($i=1,2,3$) are, e.g., given in Ref. \cite{Langlois:2015cwa}.

It should be emphasized that, in this section, we have only focused on quadratic DHOST theories.
However, in principle, we can consider DHOST theories with higher-order derivative terms (such as cubic polynomials in $D_a\phi_b$) or higher-order derivatives (such as polynomials in $D_aD_b\phi_c$).
In fact, there exist, e.g., cubic DHOST theories~\cite{BenAchour:2016fzp} having third-order derivative terms.
Also, in Ref.~\cite{Takahashi:2022mew}, the authors suggest the existence of DHOST theories including higher-order derivatives (so-called generalized disformal Horndeski theories).

\section{Degenerate vector-tensor theories in the metric formalism}
In this section, we will omit the $g$ sign of the curvature tensor for simplicity as in the previous section.

\subsection{Generalized Proca theory}
The generalized Proca Lagrangian is given by~\cite{Heisenberg:2014rta}
\begin{align}
    L=\sum_{i=2}^{6} \mathcal{L}_i,
\end{align}
where
\begin{align}
	\mathcal{L}_{2} &\coloneqq G_{2}(A_{a}, F_{a b}, \tilde{F}_{a b}),\\
	\mathcal{L}_{3} &\coloneqq G_{3}(X) D^{a} A_{a}, \\
	\mathcal{L}_{4} &\coloneqq G_{4}(X) R+G_{4, X}\left[\left(D^{a} A_{a}\right)^{2}-D_{a} A_{b} D^{b} A^{a}\right], \\
\mathcal{L}_{5} &\coloneqq G_{5}(X) G_{a b} D^{a} A^{b}-\frac{1}{6} G_{5, X}\left[(D^a A_a)^{3}+2 D_{a} A_{b} D^{c} A^{a} D^{b} A_{c}-3(D^c A_c) D_{a} A_{b} D^{a} A^{b}\right] \notag \\
		&\quad-g_{5}(X) \tilde{F}^{a c} \tilde{F}_{c}^{b} D_{a} A_{b}, \\
		\mathcal{L}_{6} &=G_{6}(X) G^{a b c d} D_{a} A_{b} D_{c} A_{d}+\frac{1}{2}G_{6, X} \tilde{F}^{c d} \tilde{F}^{a b} D_{c} A_{a} D_{d} A_{b},
\end{align}
with $X=-g^{ab}A_aA_b/2$ and
\begin{align}
    \tilde F^{ab} \coloneqq &\; \frac{1}{2}\epsilon^{abcd} F_{cd}.
\end{align}

The generalized Proca theory has five DOFs (two tensor DOFs plus three vector DOFs).
Here the zeroth component of the vector field does not propagate (see also, e.g., Ref.~\cite{Allys:2015sht}).
Note that only $G_2$ can be functions of $F$ or $\tilde F$ as well as $A$.
Furthermore, $G_2$ can be rewritten as a function of scalar combinations up to integration by parts (see, e.g., Refs.~\cite{Fleury:2014qfa,Allys:2016jaq})
\begin{align}
    G_2(A_a, F_{ab}, \tilde F{}_{ab})=G_2(X, F^2, (A\cdot F)^2, F\cdot \tilde F).
\end{align}

Note that, due to the antisymmetry of $F_{ab}$ and $G^{abcd}$, performing the scalar limit $A_a\to \partial_a\phi$ on the Generalized Proca theory leads to the (shift-symmetric) Horndeski theory.
Also, as in the Horndeski theory, the Generalized Proca theory can be expressed in terms of epsilon tensor, e.g., the term of $G_4$ can be rewritten as
\begin{align}
    \mathcal{L}_4&=G_4 R+G_{4X}[(D^aA_a)^2-D_aA_b D^bA^a]\\
    &=H_4 G^{ab}A_aA_b+\frac{1}{2}H_{4X}\epsilon^{abcd}\epsilon^{a'b'c'}{}_dA_aA_{a'}D_{b}A_{b'}D_{c}A_{c'}+(\text{total derivative}),
\end{align}
where
\begin{align}
    G_4=H_4X.
\end{align}

\subsection{Degenerate vector-tensor theories}\label{sec:EVT}
In this section, we will review ``extended vector-tensor theories~\cite{Kimura:2016rzw},'' which are the framework of the degenerate vector-tensor theories whose Lagrangian is up to polynomial in $D_aA_b$ as
\begin{align}
    L=f(X)R+G_2(X)+G_3(X)D^a A_a+C^{abcd}D_a A_b D_c A_d,
\end{align}
where
\begin{align}
	C^{a b c d} \coloneqq &\;\alpha_{1}(X) g^{a(c} g^{d) b}+\alpha_{2}(X) g^{a b} g^{c d}+\frac{1}{2} \alpha_{3}(X) \left(A^{a} A^{b} g^{c d}+A^{c} A^{d} g^{a b}\right) \notag\\
	&\qquad+\frac{1}{2} \alpha_{4}(X)\left(A^{a} A^{(c} g^{d) b}+A^{b} A^{(c} g^{d) a}\right)+\alpha_{5}(X) A^{a} A^{b} A^{c} A^{d}+\alpha_{6}(X) g^{a[c} g^{d]} b \notag\\
	&\qquad+\frac{1}{2} \alpha_{7}(X)\left(A^{a} A^{[c} g^{d] b}-A^{b} A^{[c} g^{d] a}\right)+\frac{1}{4} \alpha_{8}(X)\left(A^{a} A^{c} g^{b d}-A^{b} A^{d} g^{a c}\right)+\frac{1}{2} \alpha_{9}(X) \epsilon^{a b c d}.
\end{align}
Here, the generalized Proca theory is the special case with $\alpha_1=-\alpha_2=-f_X$ and $\alpha_3=\alpha_4=\alpha_5=\alpha_8=0$.\footnote{%
Strictly speaking, it is attributed to a subclass of generalized Proca theory up to quadratic polynomial in $D_aA_b$.
Note that this is not only a statement about $\mathcal{L}_5$ and $\mathcal{L}_6$, but also about $\mathcal{L}_2$.
Now $G_2$ should no longer be an arbitrary function of $A$, $F$, and $\tilde F$, but rather quadratic polynomial in $F$ and $\tilde F$, i.e.,
\begin{align}
    G_2(A,F,\tilde F)=G_2(X,F^2,(A\cdot F)^2, F\cdot \tilde F) \quad\Rightarrow\quad G_2(X)+\alpha_6(X) F^2+\alpha_7(X)[A\cdot F]^2+\alpha_9(X) F\cdot \tilde F.
\end{align}
}
Note that $\alpha_j$ ($j=6,7,8,9$) represents the components of the pure vector, which automatically vanishes when we perform scalar limit: $A_a\to\phi_a$.

The degeneracy conditions for this Lagrangian are given by
\begin{align}
    D_0(X)=D_1(X)=D_2(X)=0,
\end{align}
where the details are discussed in Ref.~\cite{Kimura:2016rzw}.
Here, the explicit form of $D_0$ is expressed as
\begin{align}
	D_{0}(X)=\frac{1}{16}\left(\alpha_{1}+\alpha_{2}\right) d_0\left(f, \alpha_{1}, \alpha_{2}, \alpha_{4}, \alpha_{8}, \beta\right),
\end{align}
where
\begin{align}
	d_0\left(f, \alpha_{1}, \alpha_{2}, \alpha_{4}, \alpha_{8}, \beta\right)\coloneqq &\; 8 f^{2}\left(2 \alpha_{1}-\beta-2\left(\alpha_{4}-\alpha_{8}\right) X\right)+32 X^{2} f_{X}^{2}\left(2 \alpha_{1}-\beta-2\left(\alpha_{4}+\alpha_{8}\right) X\right) \notag\\
	&-8X f\left(2 \alpha_{1} \beta-2 f_{Y}\left(2 \alpha_{1}+\beta-2\alpha_{4} X\right)-12 f_{X}^{2}+\alpha_{8}^{2} X^{2}-2 \alpha_{4} \beta X\right),
\end{align}
and
\begin{align}
	\beta=-2\alpha_6+2\alpha_7 X.
\end{align}

The explicit form of $D_1$ is given by
\begin{align}
	D_{1}(X)=\mathcal{W}_{1} \alpha_{8}^{2}+\mathcal{W}_{2} \alpha_{8}+\mathcal{W}_{3},
\end{align}
where
\begin{align}
	\mathcal{W}_{1}(X)\coloneqq &\; \frac{1}{2}\left(\alpha_{1}+\alpha_{2}\right) X^{2} f-\frac{1}{4} X\left(f+2\alpha_{1} X\right)\left( f-\left(\alpha_{1}+3 \alpha_{2}\right) X\right), \\
	\mathcal{W}_{2}(X)\coloneqq &\; \frac{1}{2}\left(\alpha_{1}+\alpha_{2}\right)\left[2\left(4 X^{2} f_{X}^{2}-f^{2}\right)-X f\left(\alpha_{1}+\alpha_{2}-2X\left(\alpha_{3}+\alpha_{4}\right)+4\alpha_{5} X^{2}\right)\right] \notag\\
		&+\frac{1}{2} X\left( \alpha_{2}- f_{X}-\alpha_{3} X\right)\left[8 \alpha_{1} X f_{X}+f\left(2\alpha_{2}+3 f_{X}-\alpha_{3} X\right)\right], \\
	\mathcal{W}_{3}(X)\coloneqq &\; \left(\alpha_{1}+\alpha_{2}\right)\Big[-\alpha_{5}\left(2\alpha_{4} X^{3} f+X f\left(-4 f-2 \alpha_{1} X+\beta X\right)\right)+\alpha_{4}^{2} X^{2} f\notag\\
		&\qquad-\alpha_{4} X\left(2 \beta f-X\left(-3 \alpha_{1} \beta-2 \alpha_{1} f_{X}-8 f_{X}^{2}\right)\right) \notag\\
		&\qquad-\left(\alpha_{1}+ f_{X}\right)\left(X\left(-2\left(\alpha_{1}+\beta\right) f_{X}-3 \alpha_{1} \beta\right)+ f\left(\alpha_{1}-\beta+3 f_{X}\right)\right)\Big] \notag\\
		&- 2\alpha_{1} X\left(\alpha_{1} \beta\left( \alpha_{1}-\alpha_{4} X\right)+ f_{X}\left(2 \alpha_{1} \beta- f_{X}\left(-2\alpha_{4} X-\beta\right)+2\alpha_{3} \alpha_{4} X^{2}+\alpha_{3} \beta X-2 \alpha_{1} \alpha_{3} X\right)\right) \notag\\
		&+\frac{1}{4} f\Big[- f_{X}\left(-12 \alpha_{1}^{2}-3 f_{X}\left(2 \alpha_{1}-\beta-2\alpha_{4} X\right)+2 \alpha_{1}\left(3 \beta-2\left(5 \alpha_{3}+\alpha_{4}\right) X\right)-2\alpha_{3} X\left(-2\alpha_{4} X-\beta\right)\right)\notag\\
		&\qquad+\beta\left(-3 \alpha_{1}^{2}+\alpha_{3}^{2} X^{2}+2\left(3 \alpha_{3}+4 \alpha_{4}\right) \alpha_{1} X\right)+2\left( \alpha_{1}+\alpha_{3} X\right)\left(3 \alpha_{1}-\alpha_{3} X\right)\left( \alpha_{1}-\alpha_{4} X\right)\Big] \notag\\
		&+\frac{1}{2}\left(\alpha_{3}+\alpha_{4}\right) f^{2}\left(-2 \alpha_{1}+\beta-2\alpha_{3} X\right).
\end{align}

Sometimes, it is more prospective to compute the specific linear combination of $D_2$ and $D_1$ than to compute the explicit form of $D_2$ itself.
Therefore, we would like to use the following combination as the counterpart of $D_2$ itself:
\begin{align}
	d_2\coloneqq&\;  D_{1}(X)+2X D_{2}(X)\\
    =&\;-\frac{D_{0}(X)}{2X}+\frac{1}{8X}\left(f+2\alpha_{1} X\right)\left[2 \alpha_{1}-2\left(\alpha_{4}+\alpha_{8}\right) X-\beta\right]\times \notag\\
	&\quad \Big[2\left(\alpha_{1}+\alpha_{2}-2\left(\alpha_{3}+\alpha_{4}\right) X+4\alpha_{5} X^{2}\right)\left( f-\left(\alpha_{1}+3 \alpha_{2}\right) X\right)+3 X\left( \alpha_{2}- f_{X}-\alpha_{3} X\right)^{2}\Big].
\end{align}
Here, recall that the notation $X\coloneqq -A^a A_a/2$ was used.

\section{Explicit form of Sec.~\ref{sec:VTInPalatini}}
\label{app:ExplicitFormPalatini}
In this section, we show the explicit form of the functions introduced in Sec.~\ref{sec:VTInPalatini}.
In Sec.~\ref{sec:VTInPalatini}, we have studied vector-tensor theories in the Palatini formalism.
Then, the explicit forms of the coefficient functions of disformation tensor [see Eq.~\eqref{kappa-soln}] are given by\footnote{%
The disformation tensor is defined as the distortion tensor for the torsionless connection.
Written explicitly, $\kappa^a{}_{bc}\Big|_{T^a{}_{bc}=0}$.}
\begin{align}
	C_0&=C_{0,X}+C_{0,Y}Y+C_{0,Z}Z, \\
	C_1&=C_{1,X}+C_{1,Y}Y+C_{1,Z}Z, \\
	C_2&=C_{2,X}+C_{2,Y}Y+C_{2,Z}Z, \\
	C_3&=
    \frac{H_2+H_3}{2 \left(G_4-2 \left(H_2+H_3\right) X\right)}, \\
    C_4&=-C_3,\\
    C_5&=-C_3,\\
	C_6&=0,\\
	C_{7}&= -\frac{\left(H_2+H_3\right) \left(G_4 \left(3 G_{4 X}+4
   \left(H_2+H_3\right)\right)-6 \left(H_2+H_3\right) X G_{4
   X}\right)}{G_4 \left(G_4-2 \left(H_2+H_3\right) X\right) \left(3
   G_4+2 \left(H_2+H_3\right) X\right)}, \\
	C_{8}&=0, \\
	C_{9}&= -\frac{4 \left(H_2+H_3\right){}^2}{\left(G_4-2 \left(H_2+H_3\right)
   X\right) \left(3 G_4+2 \left(H_2+H_3\right) X\right)}, \\
	C_{10}&= -\frac{3 G_{4 X}+H_2+H_3}{6 G_4+4 \left(H_2+H_3\right) X}, \\
	C_{11}&= \frac{3 G_4 \left(G_{4 X}+H_2+H_3\right)-4 \left(H_2+H_3\right) X G_{4
   X}}{2 G_4 \left(3 G_4+2 \left(H_2+H_3\right) X\right)}, \\
	C_{12}&=-\frac{ \left(H_2+H_3\right)}{6 G_4+4 \left(H_2+H_3\right) X}, \\
    C_{13}&=-3C_{12},
\end{align}
where $C_1=C_1(X,Y,Z)$, $C_i=C_i(X)$ ($i=2,3,\dots13$), and
{\tiny
\begin{align}
	C_{0,X}&= \frac{2 G_3 \left(H_2+H_3\right)}{4 G_4 \left(-2 H_1+H_2+H_3\right)
   X+3 G_4^2-8 \left(H_2+H_3\right) \left(4 H_1+H_2+H_3\right) X^2}, \\
	C_{0,Y}&= -\frac{\left(H_2+H_3\right) \left(4 \left(H_2+H_3\right) \left(4
   H_1+H_2+H_3\right) X-G_4 \left(4 H_1+3
   \left(H_2+H_3\right)\right)\right)}{\left(G_4-2
   \left(H_2+H_3\right) X\right) \left(4 G_4 \left(-2
   H_1+H_2+H_3\right) X+3 G_4^2-8 \left(H_2+H_3\right) \left(4
   H_1+H_2+H_3\right) X^2\right)}, \\
	C_{0,Z}&= -\left[ G_4
   \left(G_4-2 \left(H_2+H_3\right) X\right) \left(3 G_4+2
   \left(H_2+H_3\right) X\right) \left(4 G_4 \left(-2
   H_1+H_2+H_3\right) X+3 G_4^2-8 \left(H_2+H_3\right) \left(4
   H_1+H_2+H_3\right) X^2\right)\right]^{-1}\times\notag \\
   &\qquad
   \left(H_2+H_3\right)\Big[
    16 \left(H_2+H_3\right){}^2 \left(4
   H_1+H_2+H_3\right) X^2 G_{4 X}+G_4^2 \left(3 \left(4 H_1+3
   \left(H_2+H_3\right)\right) G_{4 X}+4 \left(H_2+H_3\right) \left(4
   H_1+5 \left(H_2+H_3\right)\right)\right)\notag\\
   &\qquad\qquad
   -2 G_4 \left(H_2+H_3\right)
   X \left(\left(28 H_1+13 \left(H_2+H_3\right)\right) G_{4 X}+12
   \left(H_2+H_3\right) \left(4 H_1+H_2+H_3\right)\right)
   \Big], \\
   C_{1,X}&= \frac{2 G_3 \left(H_2+H_3\right) X}{4 G_4 \left(-2 H_1+H_2+H_3\right)
   X+3 G_4^2-8 \left(H_2+H_3\right) \left(4 H_1+H_2+H_3\right) X^2},\\
	C_{1,Y}&= \frac{G_4 \left(H_2+H_3\right) \left(2 \left(8 H_1+H_2+H_3\right) X-3
   G_4\right)}{2 \left(G_4-2 \left(H_2+H_3\right) X\right) \left(4 G_4
   \left(-2 H_1+H_2+H_3\right) X+3 G_4^2-8 \left(H_2+H_3\right)
   \left(4 H_1+H_2+H_3\right) X^2\right)}, \\
	C_{1,Z}&= -\left[ G_4
   \left(G_4-2 \left(H_2+H_3\right) X\right) \left(3 G_4+2
   \left(H_2+H_3\right) X\right) \left(4 G_4 \left(-2
   H_1+H_2+H_3\right) X+3 G_4^2-8 \left(H_2+H_3\right) \left(4
   H_1+H_2+H_3\right) X^2\right) \right]^{-1}\times\notag \\
   &\qquad \left(H_2+H_3\right) X \Big[
   16 \left(H_2+H_3\right){}^2
   \left(4 H_1+H_2+H_3\right) X^2 G_{4 X}+G_4^2 \left(3 \left(4 H_1+3
   \left(H_2+H_3\right)\right) G_{4 X}+4 \left(H_2+H_3\right) \left(4
   H_1+5 \left(H_2+H_3\right)\right)\right)\notag\\
   &\qquad\qquad-2 G_4 \left(H_2+H_3\right)
   X \left(\left(28 H_1+13 \left(H_2+H_3\right)\right) G_{4 X}+12
   \left(H_2+H_3\right) \left(4 H_1+H_2+H_3\right)\right)
   \Big], \\
   C_{2,X}&= \frac{G_3 \left(G_4+2 \left(H_2+H_3\right) X\right)}{4 G_4 \left(-2
   H_1+H_2+H_3\right) X+3 G_4^2-8 \left(H_2+H_3\right) \left(4
   H_1+H_2+H_3\right) X^2}, \\
	C_{2,Y}&= \frac{\left(G_4+2 \left(H_2+H_3\right) X\right) \left(G_4 \left(4
   H_1+3 \left(H_2+H_3\right)\right)-4 \left(H_2+H_3\right) \left(4
   H_1+H_2+H_3\right) X\right)}{2 \left(G_4-2 \left(H_2+H_3\right)
   X\right) \left(4 G_4 \left(-2 H_1+H_2+H_3\right) X+3 G_4^2-8
   \left(H_2+H_3\right) \left(4 H_1+H_2+H_3\right) X^2\right)}, \\
	C_{2,Z}&= \left[G_4 \left(G_4-2
   \left(H_2+H_3\right) X\right) \left(3 G_4+2 \left(H_2+H_3\right)
   X\right) \left(4 G_4 \left(-2 H_1+H_2+H_3\right) X+3 G_4^2-8
   \left(H_2+H_3\right) \left(4 H_1+H_2+H_3\right) X^2\right)\right]^{-1}\times
   \notag \\
   &\qquad
   \Big[-2 G_4 \left(H_2+H_3\right){}^3 X^2 \left(3 G_{4 X}+4 \left(4
   H_1+H_2+H_3\right)\right)+8 \left(H_2+H_3\right){}^3 \left(4
   H_1+H_2+H_3\right) X^3 G_{4 X}\notag\\
   &\qquad\qquad+G_4^3 \left(-6 H_1 G_{4 X}-2 \left(4
   H_1-H_2-H_3\right) \left(H_2+H_3\right)\right)+G_4^2
   \left(H_2+H_3\right) X \left(\left(4 H_1+H_2+H_3\right) G_{4 X}+8
   \left(H_2+H_3\right){}^2\right)\Big],
\end{align}
}
with $C_{j,k}=C_{j,k}(X)$ ($j=0,1,2$, $k=X,Y,Z$), $Y=D ^aA_a$, and $Z=A^aA^b D_aA_b$.

After integrating out the connection, we have obtained the Lagrangian for a vector-tensor theory in the metric formalism, and the explicit forms of its coefficient functions [see Eq.~\eqref{eq:IntegrateOutInVT}] are given by
\begin{align}
    g_2&= \frac{2 G_3^2 X \left(G_4+4 \left(H_2+H_3\right) X\right)}{4 G_4
   \left(-2 H_1+H_2+H_3\right) X+3 G_4^2-8 \left(H_2+H_3\right)
   \left(4 H_1+H_2+H_3\right) X^2}+G_2, \\
    g_3&= \frac{G_3 G_4 \left(3 G_4+4 \left(H_2+H_3\right) X\right)}{4 G_4
   \left(-2 H_1+H_2+H_3\right) X+3 G_4^2-8 \left(H_2+H_3\right)
   \left(4 H_1+H_2+H_3\right) X^2}, \\
    h_3&= -\frac{G_3 G_4 \left(3 G_{4 X}+4 \left(H_2+H_3\right)\right)}{4 G_4
   \left(-2 H_1+H_2+H_3\right) X+3 G_4^2-8 \left(H_2+H_3\right)
   \left(4 H_1+H_2+H_3\right) X^2}, \\
    g_4&= G_4,
\end{align}
and
{\tiny
\begin{align}
    \alpha_1&= \frac{G_4 H_{23}}{G_4-2 H_{23} X}, \\
    \alpha_2&= \frac{G_4 \left(-G_4 H_{23} \left(2 H_1+3 H_{23}\right) X+3 G_4^2 H_1+2 H_{23}^2 \left(4 H_1+H_{23}\right)
    X^2\right)}{\left(G_4-2 H_{23} X\right) \left(4 G_4 \left(H_{23}-2 H_1\right) X+3 G_4^2-8 H_{23} \left(4
    H_1+H_{23}\right) X^2\right)}, \\
    \alpha_3&= \frac{-4 H_{23}^2 \left(4 H_1+H_{23}\right) X^2 G_{4 X}+G_4^2 \left(-\left(3 \left(2 H_1+H_{23}\right) G_{4
    X}+2 H_{23} \left(4 H_1+3 H_{23}\right)\right)\right)+4 G_4 H_{23} X \left(\left(5 H_1+2 H_{23}\right) G_{4
    X}+2 H_{23} \left(4 H_1+H_{23}\right)\right)}{\left(G_4-2 H_{23} X\right) \left(4 G_4 \left(H_{23}-2
    H_1\right) X+3 G_4^2-8 H_{23} \left(4 H_1+H_{23}\right) X^2\right)}, \\
    \alpha_4&= \frac{6 H_{23} \left(2 G_4-3 X G_{4 X}\right) G_{4 X}+8 H_{23}^2 \left(2 G_4-3 X G_{4 X}\right)+9 G_4 G_{4
    X}^2}{2 \left(G_4-2 H_{23} X\right) \left(3 G_4+2 H_{23} X\right)}, \\
    \alpha_5&= \left(2 G_4 \left(G_4-2 H_{23} X\right) \left(3 G_4+2 H_{23} X\right) \left(4 G_4 \left(H_{23}-2 H_1\right) X+3
    G_4^2-8 H_{23} \left(4 H_1+H_{23}\right) X^2\right)\right)^{-1}\times\notag\\
    &\qquad \Big[8 G_4 H_{23}^2 X^2 G_{4 X} \left(3 \left(5 H_1+2 H_{23}\right) G_{4 X}+8 H_{23} \left(4
    H_1+H_{23}\right)\right)-24 H_{23}^3 \left(4 H_1+H_{23}\right) X^3 G_{4 X}^2
    \notag\\
    &\qquad\qquad+G_4^3 \left(9 \left(2
    H_1+H_{23}\right) G_{4 X}^2+12 H_{23} \left(4 H_1+3 H_{23}\right) G_{4 X}+8 H_{23}^2 \left(4 H_1+5
    H_{23}\right)\right)\notag\\
    &\qquad\qquad-4 G_4^2 H_{23} X \left(9 \left(2 H_1+H_{23}\right) G_{4 X}^2+2 H_{23} \left(28 H_1+13
    H_{23}\right) G_{4 X}+12 H_{23}^2 \left(4 H_1+H_{23}\right)\right)\Big],\\
    \alpha_6&= H_3-H_2=-h_{23}, \\
    \alpha_7&= \frac{9 G_{4 X}^2}{6 G_4+4 H_{23} X}, \\
    \alpha_8&= -\frac{3 G_{4 X} \left(3 G_{4 X}+2 H_{23}\right)}{3 G_4+2 H_{23} X}, \\
    \alpha_9&= 0.
\end{align}
}
with $H_{23}\coloneqq H_2+H_3$ and $h_{23}\coloneqq H_2-H_3$.

Finally, let us show the explicit forms of the degeneracy conditions for this system:
{\tiny
\begin{align}
    D_0&=\frac{G_4 \left(G_4-X G_{4 X}\right){}^2 \left(G_4 H_{23} \left(H_{23}-10 H_1\right) X+3 G_4^2
    \left(H_1+H_{23}\right)-6 H_{23}^2 \left(4 H_1+H_{23}\right) X^2\right) \left(-3 G_4
    \left(h_{23}-H_{23}\right)-8 h_{23} H_{23} X\right)}{\left(G_4-2 H_{23} X\right) \left(3 G_4+2 H_{23}
    X\right) \left(4 G_4 \left(H_{23}-2 H_1\right) X+3 G_4^2-8 H_{23} \left(4 H_1+H_{23}\right) X^2\right)},\\
    D_1&=-G_4 H_{23} \left(G_4-X G_{4 X}\right){}^2\left[ 2 \left(G_4-2
    H_{23} X\right) \left(3 G_4+2 H_{23} X\right) \left(4 G_4 \left(H_{23}-2 H_1\right) X+3 G_4^2-8 H_{23}
    \left(4 H_1+H_{23}\right) X^2\right)\right]^{-1}\times\notag \\
    &\qquad\Big[-4 G_4 H_{23} X \left(h_{23} \left(4 H_{23}-37
    H_1\right)+9 H_{23} \left(4 H_1+H_{23}\right)\right)+3 G_4^2 \left(h_{23} \left(4 H_1-7 H_{23}\right)+H_{23}
    \left(H_{23}-10 H_1\right)\right)+96 h_{23} H_{23}^2 \left(4 H_1+H_{23}\right) X^2\Big],\\
    d_2&=\left[-2 X \left(2 H_{23} X-G_4\right){}^3 \left(3 G_4+2 H_{23} X\right){}^2 \left(G_4 \left(8 H_1 X-4 H_{23}
    X\right)-3 G_4^2+8 H_{23} \left(4 H_1+H_{23}\right) X^2\right){}^2\right]^{-1}\times\notag\\
    &\qquad \left[\frac{1}{2} G_4 \left(G_4-2 H_{23} X\right){}^2 \left(3 G_4+2 H_{23} X\right)\right]\times\notag\\
    &\qquad \Bigg[ 2 \left(G_4-X G_{4
    X}\right){}^2 \left(4 G_4 \left(H_{23}-2 H_1\right) X+3 G_4^2-8 H_{23} \left(4 H_1+H_{23}\right) X^2\right)\times\notag\\
    &\qquad\qquad\left(G_4 H_{23} \left(H_{23}-10 H_1\right) X+3 G_4^2 \left(H_1+H_{23}\right)-6 H_{23}^2 \left(4
    H_1+H_{23}\right) X^2\right) \left(3 G_4 \left(h_{23}-H_{23}\right)+8 h_{23} H_{23} X\right)\notag\\
    &\quad\qquad+ 3 G_4 \left(3 G_4 \left(h_{23}-H_{23}\right)+2 h_{23}
    H_{23} X\right) \times\notag \\
    &\qquad\qquad \Big\{ 36
    H_{23}^2 \left(4 H_1+H_{23}\right){}^2 X^5 G_{4 X}^2+4 G_4 H_{23} \left(4 H_1+H_{23}\right) X^4 G_{4 X}
    \left(\left(46 H_1+H_{23}\right) G_{4 X}+9 H_{23} \left(4 H_1+H_{23}\right)\right)\notag\\
    &\qquad\qquad\qquad+G_4^2 X^3 \left(\left(212
    H_1^2-164 H_{23} H_1-43 H_{23}^2\right) G_{4 X}^2+2 \left(44 H_1-19 H_{23}\right) H_{23} \left(4
    H_1+H_{23}\right) G_{4 X}+9 H_{23}^2 \left(4 H_1+H_{23}\right){}^2\right)\notag\\
    &\qquad\qquad\qquad+2 G_4^3 X^2 \left(-45 H_1 G_{4
    X}^2+\left(26 H_1^2-47 H_{23} H_1-H_{23}^2\right) G_{4 X}+H_{23} \left(4 H_1+H_{23}\right) \left(17 H_1+8
    H_{23}\right)\right)\notag\\
    &\qquad\qquad\qquad+G_4^4 X \left(-6 \left(H_1-2 H_{23}\right) G_{4 X}+9 G_{4 X}^2+25 H_1^2-8 H_{23}^2+8
    H_1 H_{23}\right)-6 G_4^5 \left(H_1+H_{23}\right)\Big\} \Bigg],
\end{align}
}
with $d_2\coloneqq D_1+2XD_2$.

\section{Explicit forms of Sec.~\ref{sec:VTInMA}}
In this section, we show the explicit form of the functions introduced in Sec.~\ref{sec:VTInMA}.
In Sec.~\ref{sec:VTInMA}, we have studied projective-invariant vector-tensor theories in the metric affine formalism.
Then, to obtain the projective-invariant Lagrangian, we have considered the following two options.

\subsection{Contracting the epsilon tensor}\label{app:ExplicitFormEpsilon}
In Sec.~\ref{sec:epsilonVT}, we used the epsilon tensor to get a projective invariant Lagrangian.
Then, the explicit forms of the coefficient functions of the distortion tensor [see Eq.~\eqref{gen:kappa}] are given by
{\tiny
\begin{align}
	C_0&=-\frac{G_3 \left(4 F X+G_4\right)+\left(H_1+H_2\right) \left(G_4 Y-Z G_{4 X}\right)}{G_4 \left(4 X\left(F+\left(H_1+H_2\right) X\right)+G_4\right)},\\
	C_1&=-C_2=\frac{Z G_{4 X} \left(F+\left(H_1+H_2\right) X\right)-G_3 G_4 X}{G_4 \left(4 X\left(F+\left(H_1+H_2\right) X\right)+G_4\right)},\\
	C_3&=\frac{2 X \left(2 G_4 \left(F^2+2 F \left(H_1-H_2\right) X+2 \left(H_1^2-H_2^2\right) X^2\right)+8F^2 X \left(F+\left(H_1+H_2\right) X\right)-G_4^2 H_2\right)}{G_4 \left(16 X^2 \left(F^2+2 FH_2 X+\left(H_2^2-H_1^2\right) X^2\right)+8 G_4 X \left(F+H_2 X\right)+G_4^2\right)},\\
	C_4&=-\frac{2 \left(8 F^2 G_4 X^2 \left(4 F+3 \left(H_1+H_2\right) X\right)+2 G_4^2 X \left(5 F^2+2 FH_1 X+2 \left(H_1^2-H_2^2\right) X^2\right)+32 F^3 X^3 \left(F+\left(H_1+H_2\right)X\right)+G_4^3 \left(F-H_2 X\right)\right)}{G_4^2 \left(16 X^2 \left(F^2+2 F H_2X+\left(H_2^2-H_1^2\right) X^2\right)+8 G_4 X \left(F+H_2 X\right)+G_4^2\right)},\\
	C_5&=\frac{2 \left(2 G_4 X \left(3 F^2-2 F \left(H_1-3 H_2\right) X+2 \left(H_2^2-H_1^2\right)X^2\right)+8 F X^2 \left(F^2-F \left(H_1-3 H_2\right) X+2 \left(H_2^2-H_1^2\right)X^2\right)+G_4^2 \left(F+H_2 X\right)\right)}{G_4 \left(16 X^2 \left(F^2+2 F H_2X+\left(H_2^2-H_1^2\right) X^2\right)+8 G_4 X \left(F+H_2 X\right)+G_4^2\right)},\\
	C_6&=-\frac{2 \left(8 F^2 X^2 \left(F+\left(H_1+H_2\right) X\right)+2 F G_4 X \left(3 F+2\left(H_1+H_2\right) X\right)+G_4^2 \left(F+H_1 X\right)\right)}{G_4 \left(16 X^2 \left(F^2+2 FH_2 X+\left(H_2^2-H_1^2\right) X^2\right)+8 G_4 X \left(F+H_2 X\right)+G_4^2\right)},\\
	C_7&=\frac{2 \left(4 F X+G_4\right) \left(8 F^2 X^2 \left(F+\left(H_1+H_2\right) X\right)+2 F G_4 X\left(3 F+2 \left(H_1+H_2\right) X\right)+G_4^2 \left(F+H_1 X\right)\right)}{G_4^2 \left(16 X^2\left(F^2+2 F H_2 X+\left(H_2^2-H_1^2\right) X^2\right)+8 G_4 X \left(F+H_2X\right)+G_4^2\right)},\\
	C_8&=\frac{2 X \left(-8 F X \left(F^2-F \left(H_1-3 H_2\right) X+2 \left(H_2^2-H_1^2\right)
	X^2\right)-2 F G_4 \left(F+2 \left(H_2-H_1\right) X\right)+G_4^2 H_1\right)}{G_4 \left(16 X^2
	\left(F^2+2 F H_2 X+\left(H_2^2-H_1^2\right) X^2\right)+8 G_4 X \left(F+H_2
	X\right)+G_4^2\right)},\\
	C_9&=\left[G_4^2 \left(8 G_4 X \left(F+H_2 X\right)+16 X^2\left(\left(F+H_2 X\right){}^2-H_1^2 X^2\right)+G_4^2\right)\right]^{-1}\times\notag\\
		&\qquad \Bigg[-8 G_4 X \left(F \left(F^2+G_{4 X} \left(F+H_2 X\right)+4 F H_1 X-4 F H_2 X+4 H_1^2 X^2-4H_2^2 X^2\right)-4 X F_X \left(\left(F+H_2 X\right){}^2-H_1^2 X^2\right)\right)\notag\\
		&\quad\qquad-16 F X^2\left(F+H_1 X+H_2 X\right) \left(2 F^2+G_{4 X} \left(F-H_1 X+H_2 X\right)\right)+G_4^3 \left(2F_X-H_1+H_2\right)\notag\\
		&\quad\qquad+G_4^2 \left(-F G_{4 X}+16 X F_X \left(F+H_2 X\right)+4 X \left(-2 F H_1+3 FH_2+H_1^2 (-X)+H_2^2 X\right)\right)\Bigg],\\
	C_{10}&=\left[G_4^2 \left(8 G_4 X \left(F+H_2 X\right)+16 X^2 \left(\left(F+H_2X\right){}^2-H_1^2 X^2\right)+G_4^2\right)\right]^{-1}\times \notag\\
		&\qquad \Bigg[8 G_4 X \left(F \left(F^2+G_{4 X} \left(F+H_2 X\right)+2 F H_1 X-2 F H_2 X+2 H_1^2 X^2-2H_2^2 X^2\right)-4 X F_X \left(\left(F+H_2 X\right){}^2-H_1^2 X^2\right)\right)\notag\\
		&\quad\qquad+16 F X^2\left(F+H_1 X+H_2 X\right) \left(2 F^2+G_{4 X} \left(F-H_1 X+H_2 X\right)\right)+G_4^3 \left(-2F_X+H_1+H_2\right)\notag\\
		&\quad\qquad+G_4^2 \left(F G_{4 X}-16 X F_X \left(F+H_2 X\right)+4 X \left(F H_1+H_1^2(-X)+H_2^2 X\right)\right)\Bigg],\\
	C_{11}&=\frac{\left(H_1-H_2\right) \left(4 F X+G_4\right)}{G_4 \left(4 X \left(F+\left(H_2-H_1\right)X\right)+G_4\right)},\\
	C_{12}&=\frac{8 F G_4 X \left(3 F^2+4 F H_1 X+2 \left(H_1^2-H_2^2\right) X^2\right)+4 G_4^2 \left(F^2+F\left(3 H_1-2 H_2\right) X+\left(H_1^2-H_2^2\right) X^2\right)+32 F^3 X^2\left(F+\left(H_1+H_2\right) X\right)+G_4^3 \left(H_1-H_2\right)}{G_4^2 \left(16 X^2\left(F^2+2 F H_2 X+\left(H_2^2-H_1^2\right) X^2\right)+8 G_4 X \left(F+H_2X\right)+G_4^2\right)},\\
	C_{13}&=\frac{-4 G_4^2 \left(F^2-F H_2 X+\left(H_1^2-H_2^2\right) X^2\right)-8 F^2 G_4 X \left(3 F+2\left(H_1+H_2\right) X\right)-32 F^3 X^2 \left(F+\left(H_1+H_2\right) X\right)+G_4^3\left(H_1+H_2\right)}{G_4^2 \left(16 X^2 \left(F^2+2 F H_2 X+\left(H_2^2-H_1^2\right)X^2\right)+8 G_4 X \left(F+H_2 X\right)+G_4^2\right)},\\
	C_{14}&=-\frac{\left(H_1-H_2\right) \left(4 F X+G_4\right)}{G_4 \left(4 X \left(F+\left(H_2-H_1\right)X\right)+G_4\right)},\\
	C_{15}&=-\frac{G_{4 X}}{2 G_4},\\
    C_{16}&=-C_{15},\\
	C_{17}&=0,\\
    C_{18}&=C_{17},
\end{align}
}
where $C_i=C_i(X)$ ($i=0,1,\dots 18$).

After integrating out the connection, we have obtained the Lagrangian for a vector-tensor theory in the metric formalism, and the explicit forms of its coefficient functions [see Eq.~\eqref{eq:IntegrateOutInVT}] are given by
\begin{align}
    g_2&= G_2-\frac{12 G_3^2 X^3}{X \left(F+4 \left(H_1+H_2\right) X\right)+G_4},\\
    g_3&= \frac{4 G_3 G_4 X}{X \left(F+4 \left(H_1+H_2\right) X\right)+G_4},\\
    h_3&= \frac{2 G_3 \left(G_4-3 X G_{4 X}\right)}{X \left(F+4 \left(H_1+H_2\right) X\right)+G_4},\\
    g_4&= F X+G_4,
\end{align}
and
{\tiny
\begin{align}
    \alpha_1&= -\alpha_2= -\frac{2 G_4 \left(F+2 H_1 X+2 H_2 X\right)+F X \left(F+4 H_1 X+4 H_2 X\right)}{2 \left(X \left(F+4 H_1 X+4 H_2
   X\right)+G_4\right)},\\
    \alpha_3&= -\frac{\left(F+2 G_{4 X}+2 X G_{5 X}\right) \left(F+4 \left(H_1+H_2\right) X\right)+2 G_4 \left(G_{5 X}-2
   \left(H_1+H_2\right)\right)}{2 \left(X \left(F+4 \left(H_1+H_2\right) X\right)+G_4\right)},\\
    \alpha_4&= \left[2 G_4^2 \left(X \left(F+4 H_1 X+4 H_2 X\right)+G_4\right)\right]^{-1}\times \notag \\
    &\qquad\Big[-F X^2 G_{4 X}^2 \left(F+4 H_1 X+4 H_2 X\right)+G_4^2 \left(\left(F+2 X G_{5 X}\right) \left(F+4 H_1 X+4
   H_2 X\right)+2 G_{4 X} \left(F+2 X G_{5 X}\right)+3 G_{4 X}^2\right)\notag\\
   &\quad\qquad+2 G_4 X G_{4 X} \left(\left(F+2 X G_{5
   X}\right) \left(F+4 H_1 X+4 H_2 X\right)+G_{4 X} \left(F+6 H_1 X+6 H_2 X\right)\right)+2 G_4^3 \left(G_{5
   X}-2 \left(H_1+H_2\right)\right)\Big],\\
    \alpha_5&= \frac{G_{4 X} \left(2 G_4 \left(\left(F+2 X G_{5 X}\right) \left(F+4 \left(H_1+H_2\right) X\right)+G_{4 X}
   \left(F+6 \left(H_1+H_2\right) X\right)\right)-F X G_{4 X} \left(F+4 \left(H_1+H_2\right) X\right)+4 G_4^2
   \left(G_{5 X}-2 \left(H_1+H_2\right)\right)\right)}{4 G_4^2 \left(X \left(F+4 \left(H_1+H_2\right)
   X\right)+G_4\right)},\\
    \alpha_6&= \frac{\left(F X+G_4\right) \left(3 F^2 G_4 X+F^3 X^2+2 G_4^2 \left(F+2 H_1 X-2 H_2 X\right)\right)}{2 G_4^2
   \left(X \left(F-4 H_1 X+4 H_2 X\right)+G_4\right)},\\
    \alpha_7&= -\frac{1}{2 G_4^2 X^2}\left[-\frac{\left(F X+G_4\right){}^4}{X \left(F-4 H_1 X+4 H_2 X\right)+G_4}+F X^3 G_{4 X}^2-6 F G_4 X^2 G_{4
   X}+3 F G_4^2 X-3 G_4 X^2 G_{4 X}^2+2 G_4^2 X^2 G_{5 X}-4 G_4 X^3 G_{4 X} G_{5 X}+G_4^3\right],\\
    \alpha_8&= \frac{G_{4 X} \left(F X G_{4 X}-G_4 \left(4 \left(F+X G_{5 X}\right)+3 G_{4 X}\right)\right)}{G_4^2},\\
    \alpha_9&= 0.
\end{align}
}
In this case, the degeneracy conditions for this system are identically satisfied:
\begin{align}
    D_0=D_1=D_2=0.
\end{align}

\subsection{Projective-invariant Lagrangian obtained in a different way}\label{app:ExplicitFormMinimal}
In Sec.~\ref{sec:minimalVT}, we utilized the projective-invariant combination to get a projective invariant Lagrangian.
Then, the explicit forms of the coefficient functions of the distortion tensor [see Eq.~\eqref{gen:kappa}] are given by
\begin{align}
    C_0&=0,\\
    C_1&=C_{1,X} (X)+C_{1,Y} (X)Y+C_{1,Z} (X)Z,\\
    C_2&=-C_1,\\
    C_3&=\frac{Q_2+Q_3}{2 \left(G_4-2 \left(Q_2+Q_3\right) X\right)},\\
    C_4&=\frac{Q_2-Q_3}{2 G_4},\\
    C_5&=-C_3,\\
    C_6&=-C_3,\\
    C_7&=-C_4,\\
    C_8&=-C_3,\\
    C_9&=\frac{G_4 \left(Q_3 \left(G_{4 X}+Q_3\right)+Q_2^2+Q_3 Q_2\right)-Q_3 \left(Q_2+Q_3\right) X \left(2 G_{4
	X}-Q_2+Q_3\right)}{2 G_4 \left(G_4-Q_3 X\right) \left(G_4-2 \left(Q_2+Q_3\right) X\right)},\\
    C_{10}&=-C_9,\\
    C_{11}&=0,\\
    C_{12}&=\frac{G_4 Q_2 \left(Q_2+2 Q_3\right)+Q_3 \left(Q_3^2-Q_2^2\right) X}{2 G_4 \left(G_4-Q_3 X\right)
	\left(G_4-2 \left(Q_2+Q_3\right) X\right)},\\
    C_{13}&=-C_{12},\\
    C_{14}&=0,\\
    C_{15}&=\frac{2 Q_3 X G_{4 X}-G_4 \left(G_{4 X}+Q_2\right)}{2 G_4 \left(G_4-Q_3 X\right)},\\
    C_{16}&=-C_{15},\\
    C_{17}&=\frac{Q_3}{2 Q_3 X-2 G_4},\\
    C_{18}&=-C_{17},
\end{align}
where $C_{i}=C_i(X,Y,Z)$ ($i=1,2$), $C_{j}=C_j(X)$ ($j=0,3,4,\dots,18$) and
\begin{align}
	C_{1,X}&=-\frac{G_3}{4 \left(G_4+\left(3 Q_1+Q_2+Q_3\right) X\right)},\\
	C_{1,Y}&=-\frac{G_4 \left(Q_1+Q_2+Q_3\right)}{2 \left(G_4-2 \left(Q_2+Q_3\right) X\right) \left(G_4+\left(3Q_1+Q_2+Q_3\right) X\right)},\\
	C_{1,Z}&=\Big[4 G_4 \left(G_4-Q_3 X\right)\left(G_4-2 \left(Q_2+Q_3\right) X\right) \left(G_4+\left(3 Q_1+Q_2+Q_3\right) X\right)\Big]^{-1}\times\notag\\
	&\qquad \Big[4 Q_3 \left(Q_2+Q_3\right) \left(3 Q_1+Q_2+Q_3\right) X^2 G_{4 X}\notag\\
	&\quad\qquad-2 G_4 X\left(\left(\left(Q_2+Q_3\right){}^2+3 Q_1 \left(Q_2+2 Q_3\right)\right) G_{4 X}+\left(Q_2+Q_3\right)\left(\left(Q_2+Q_3\right){}^2+Q_1 \left(3 Q_2+2 Q_3\right)\right)\right)\notag\\
	&\quad\qquad+G_4^2 \left(\left(3Q_1+Q_2\right) G_{4 X}+\left(Q_1-Q_2\right) \left(Q_2+Q_3\right)\right)\Big],
\end{align}
with $C_{1,k}=C_{1,k}(X)$ ($k=X,Y,Z$), $Y=D ^aA_a$, and $Z=A^aA^b D_aA_b$.

After integrating out the connection, we have obtained the Lagrangian for a vector-tensor theory in the metric formalism, and the explicit forms of its coefficient functions [see Eq.~\eqref{eq:IntegrateOutInVT}] are given by
\begin{align}
    g_2&=G_2-\frac{3 G_3^2 X}{4 \left(G_4+\left(3 Q_1+Q_2+Q_3\right) X\right)},\\
    g_3&=\frac{G_3 G_4}{G_4+\left(3 Q_1+Q_2+Q_3\right) X},\\
    h_3&=-\frac{G_3 \left(3 G_{4 X}+Q_2+Q_3\right)}{2 \left(G_4+\left(3 Q_1+Q_2+Q_3\right) X\right)},\\
    g_4&=G_4,\\
\end{align}
and
\begin{align}
    \alpha_1&=\frac{G_4 \left(Q_2+Q_3\right)}{G_4-2 \left(Q_2+Q_3\right) X},\\
    \alpha_2&=\frac{G_4 \left(G_4 Q_1-\left(Q_2+Q_3\right) \left(3 Q_1+Q_2+Q_3\right) X\right)}{\left(G_4-2
    \left(Q_2+Q_3\right) X\right) \left(G_4+\left(3 Q_1+Q_2+Q_3\right) X\right)},\\
    \alpha_3&=-\frac{2 \left(Q_2+Q_3\right){}^2}{3 G_4-6 \left(Q_2+Q_3\right) X}-\frac{\left(3 Q_1+Q_2+Q_3\right) \left(3
    G_{4 X}+Q_2+Q_3\right)}{3 \left(G_4+\left(3 Q_1+Q_2+Q_3\right) X\right)},\\
    \alpha_4&=\frac{\left(-G_{4 X}+Q_2+Q_3\right){}^2}{2 Q_3 X-2 G_4}-\frac{2 \left(Q_2+Q_3\right){}^2}{2
    \left(Q_2+Q_3\right) X-G_4}+\frac{2 G_{4 X}^2}{G_4},\\
    \alpha_5&=\frac{1}{12 G_4}\left[-\frac{8 \left(Q_2+Q_3\right){}^3}{2 \left(Q_2+Q_3\right) X-G_4}+\frac{3 Q_3 \left(-G_{4
    X}+Q_2+Q_3\right){}^2}{Q_3 X-G_4}+\frac{\left(3 Q_1+Q_2+Q_3\right) \left(3 G_{4
    X}+Q_2+Q_3\right){}^2}{G_4+\left(3 Q_1+Q_2+Q_3\right) X}\right],\\
    \alpha_6&=Q_3-Q_2,\\
    \alpha_7&=\frac{G_4 \left(G_{4 X}+Q_2-Q_3\right) \left(3 G_{4 X}-Q_2+Q_3\right)-4 Q_3 X G_{4 X}^2}{2 G_4 \left(G_4-Q_3
    X\right)},\\
    \alpha_8&=\frac{4 Q_3 X G_{4 X}^2-G_4 \left(\left(3 G_{4 X}-Q_2\right) \left(G_{4 X}+Q_2\right)+Q_3^2\right)}{G_4
    \left(G_4-Q_3 X\right)},\\
    \alpha_9&=0.
\end{align}

Explicit form of the degeneracy condition $D_0=0$ for this system is given by
\begin{align}
    D_0&=\frac{32 G_4^3 Q_3}{G_4-Q_3 X}.
\end{align}
Note that, in this case, other degeneracy conditions (i.e., $D_1=D_2=0$) are identically satisfied.

\bibliography{refs}
\bibliographystyle{JHEP}
\end{document}